\documentclass[twocolumn,showpacs,preprintnumbers,amsmath,amssymb,superscriptaddress,nofootinbib]{revtex4}

\usepackage[english]{babel}
\usepackage{amsmath}
\usepackage{graphicx}
\usepackage{dcolumn}
\usepackage{bm}
\usepackage{amscd}
\usepackage{hyperref}
\usepackage[utf8x]{inputenc}
\usepackage{textcomp}
\begin{document}
\title{Hitting and Trapping Times on Branched Structures}
\author{Elena Agliari}
\affiliation{Dipartimento di Fisica, Sapienza Università di Roma}

\author{Fabio Sartori}
\affiliation{Dipartimento di Fisica e Scienze della Terra, I-43124 Parma, Italy}

\author {Luca Cattivelli}
\affiliation{Scuola Normale Superiore, Pisa, Italy}

\author {Davide Cassi}
\affiliation{Dipartimento di Fisica e Scienze della Terra, I-43124 Parma, Italy}

\begin{abstract}
In this work we consider a simple random walk embedded in a generic branched structure and we find a close-form formula to calculate the hitting time $H\left(i,f\right)$ between two arbitrary nodes $i$ and $j$. We then use this formula to obtain the set of hitting times $\left\{ H\left(i,f\right)\right\} $ for combs and their expectation values, namely the mean-first passage time ($\mbox{MFPT}_{f})$, where the average is performed over the initial node while the final node $f$ is given, and the global mean-first passage time (GMFPT), where the average is performed over both the initial and the final node. Finally, we discuss applications in the context of reaction-diffusion problems. 

\end{abstract}

\maketitle

\section{Introduction}

Random walks (RWs) on inhomogeneous structures have been first introduced to describe anomalous diffusion \cite{Havlin,Sokolov,Metzler2}, but they actually constitute a convenient model for many real phenomena, ranging from soft matter (e.g. gels and biological structures \cite{Biology}), to condensed matter (e.g., fractures \cite{fractures} and light scattering \cite{Scattering}).

\begin{figure}
	\includegraphics[width=\linewidth]{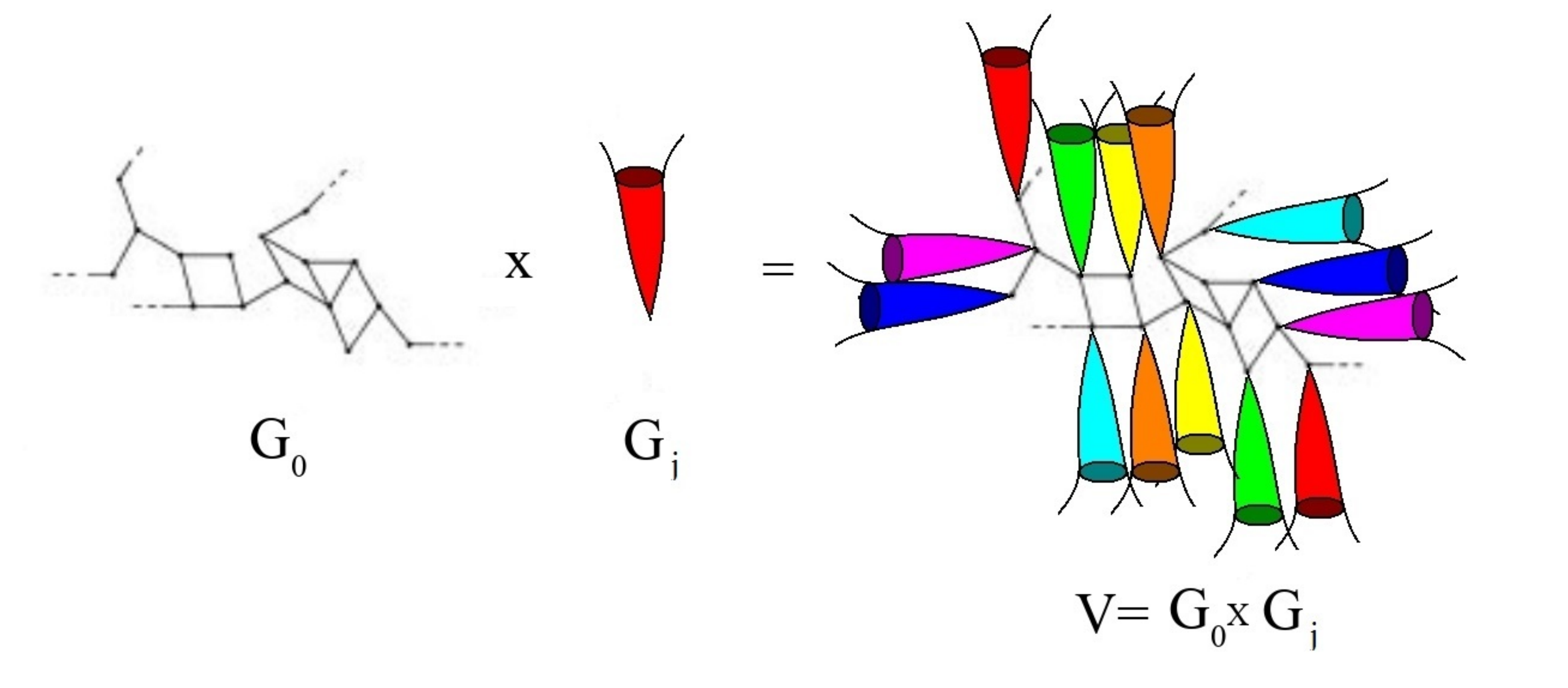} 
	\caption{Example of branched structure $\mathcal{V}$ obtained by joining to each point $j$ of a graph $\mathcal{G}_{0}$ called base (leftmost panel) a graph $\mathcal{G}_{j}$ called fiber (middle panel). \label{fig: grafo ramificato generico}}
\end{figure}

\begin{figure}
	\includegraphics[width=\linewidth]{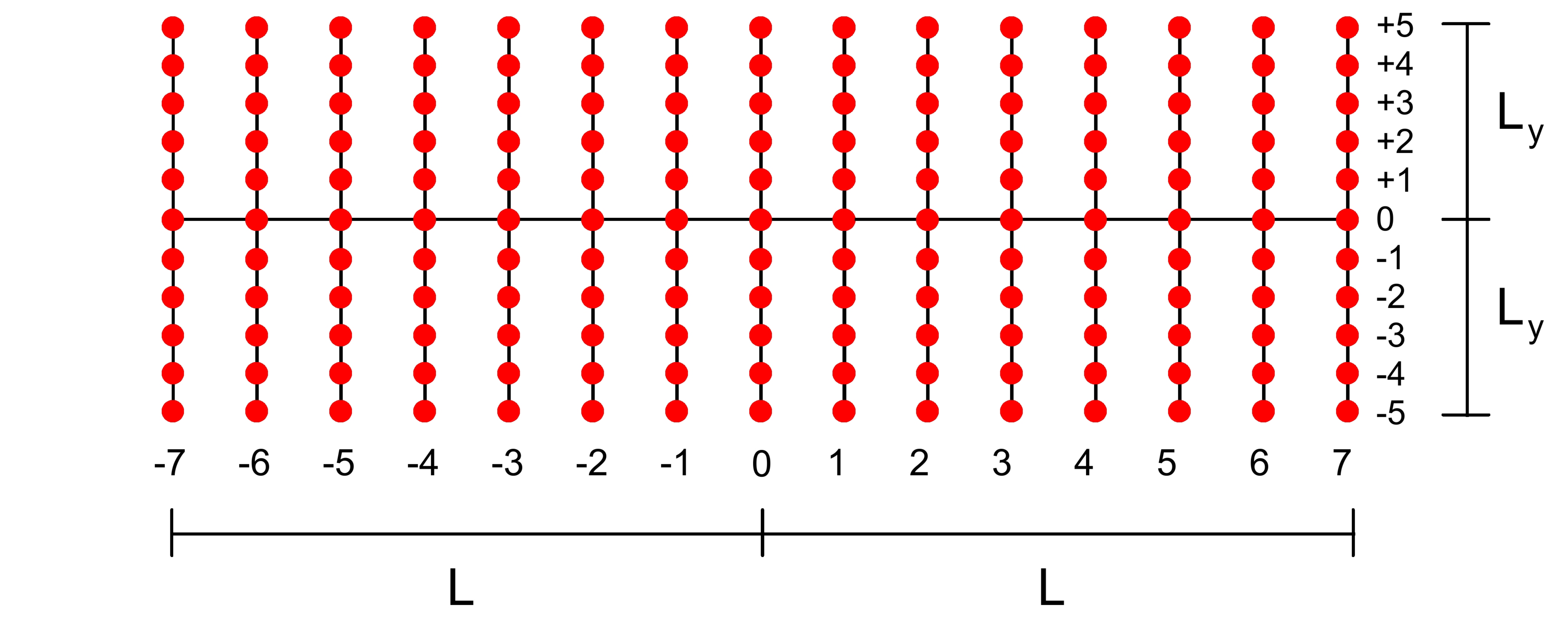}
	\includegraphics[width=\linewidth]{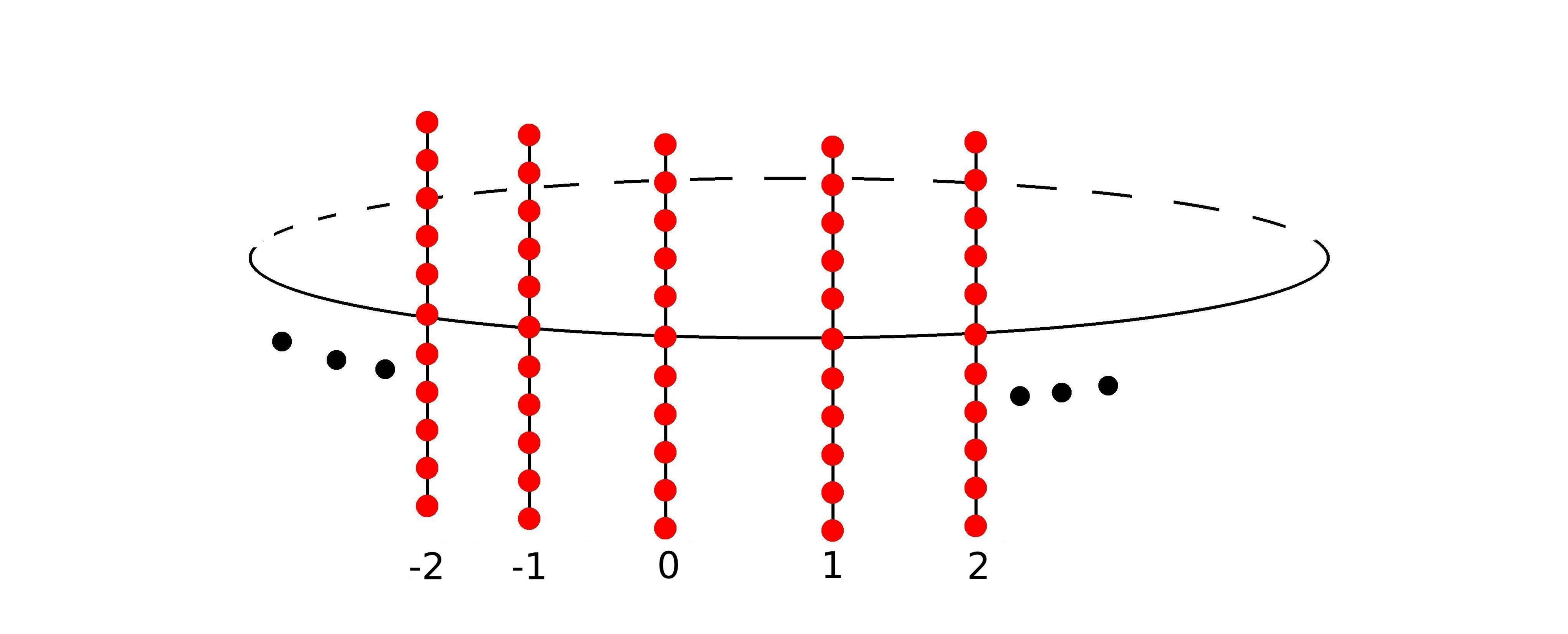}  
	\caption{Examples of $2$-dimensional comb, where the backbone is endowed with reflecting boundary conditions (upper panel) or periodic boundary conditions (lower panel). The size of the backbone is referred to as $2L+1$, while the length of teeth $L_y$ is taken to be uniform and scaling linearly with $L$, namely $L_y = \alpha L$. For instance, in the upper panel $L=7$ and $\alpha L=5$.} \label{fig: Pettini}
\end{figure}

Here we will focus on the so-called ``branched structures'', namely graphs $\mathcal{V}$ obtained by attaching to each vertex $j$ of a base graph $\mathcal{G}_{0}$ a different graph $\mathcal{G}_{j}$ called fiber (see Fig. \ref{fig: grafo ramificato generico}). 
Branched strucutres were used to describe the energy transfer in polymers \cite{27}, the transport of calcium in spiny dendrites \cite{dendy1,dendy2,dendy3}, the excitation of nano-antennas \cite{35}, the anomalous diffusion in real structure with similar topology \cite{Havlin,C-anom1,C-anom2,22,24,26}.
In particular, we will consider a specific class of branched structures called combs, namely graphs where the base graph as well the fiber ones are linear chains, in such a way that $\mathcal{G}_{j}$ is site independent and equivalent to $\mathbb{Z}$ (see Fig.~\ref{fig: Pettini}).

Diffusion on combs has been shown  to display many peculiar features ultimately stemming from their highly inhomogenous topology (see e.g.,\cite{FCP1,FCP2,FCP3,FCP4}).
Now, most of the previous results have been proven in the thermodynamic limit, namely for structures of infinite size, while real systems are intrinsically finite. In this work we aim to investigate the problem of diffusion in finite comb lattices and we will especially focus on first-passage quantities such as the hitting time $H\left(i,f\right)$ from $i$ to $f$ (i.e. the mean time for a random walker to first reach site $f$ starting from $i$), the mean first-passage time $\mbox{MFPT}_f$ to $f$ (i.e., the mean time needed to first reach the vertex $f$, averaged over the starting site), and the global mean-first passage time GMFPT (i.e., the mean time to go from a random vertex to a second random vertex).

These quantities have been extensively studied in the past, also due to the number of different applications in several research areas: pharmacokinetics \cite{Farmaco}, reaction-diffusion processes \cite{ReacDiff}, excitation transport in photosystems \cite{Montroll,Blumen2}, target search processes \cite{Predator}, disease spreading \cite{Info} and many other physical problems \cite{Redner1,Redner2,Benichou1,Benichou2,MABV-PRE2012,A-PRE2008,LowDim}. Pharmacokinetics and reaction-diffusion processes could also be affected by the occupation time, namely the mean time spent in a subset of graph's vertices \cite{Occupation1,Occupation2}.

In this work we calculate the above mentioned quantities for ($d$-dimensional) combs using the resistance method: the original graph $\mathcal{V}$ is mapped into a resistance network by replacing any link $\left(a,b\right)$ between two adjacent nodes $a$ and $b$ with a unitary resistance $R$ \cite{Snell}. Then, we use Tetali's formula \cite{Tetali} to calculate the exact value of the hitting time between $i$ and $f$ as

\begin{equation}
	H\left(i,f\right)=m R_{i,f}+\frac{1}{2}\sum_{z\in V}g\left(z\right)\left\{ R_{f,z}-R_{i,z}\right\} ,\label{eq: Tetali}
\end{equation}
where $m$ is the number of links in the graph, $g(z)$ is the coordination number of the vertex $z$, and $R_{a,b}$ is the effective resistance between the vertices $a$ and $b$.

\section{Hitting Time of Branched Structures}

In this section we will use Eq.(\ref{eq: Tetali}) to calculate the value of the hitting time for generic branched structures (see Fig.~\ref{fig: grafo ramificato generico}). 

\begin{figure}[htbp]
	\includegraphics[width=\linewidth]{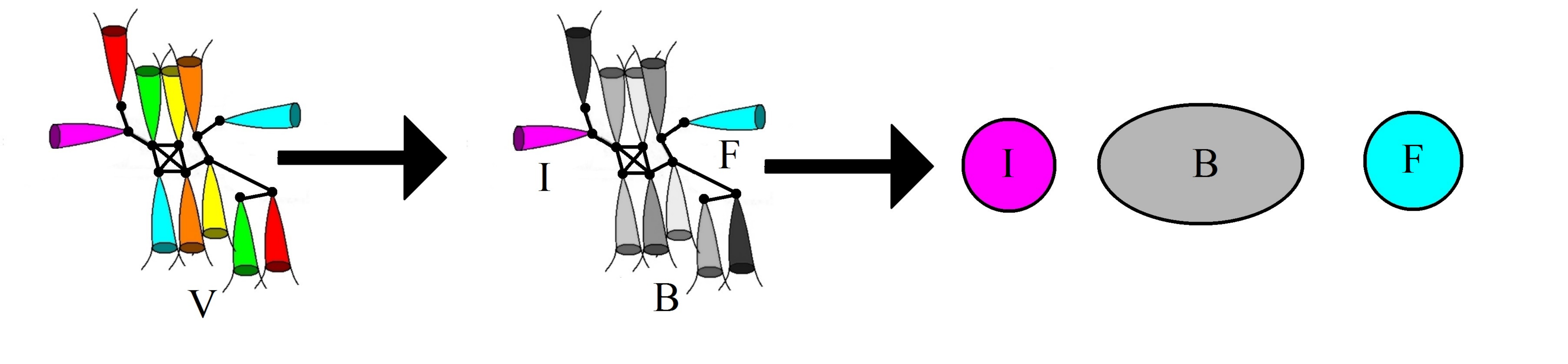} 
	\caption{\label{fig: tripartizione grafo} Schematization of the decomposition used in Eq. (\ref{eq: 2) Decomp. Sum}): where vertices of $\mathcal{V}$ are  $V = \mathcal{I}\cup\mathcal{F}\cup\mathcal{B}$. Notice that the whole set of fiber graphs $\{ \mathcal{G}_j \}$ also includes the set of nodes making up the base graph because the joining nodes, also called ``roots'', are shared.}
\end{figure}

When the starting point $i$ and the ending point $f$ belong to different fibers, we can divide vertices of $\mathcal{V}$ into three disjoint subsets, referred to as $\mathcal{I}, \mathcal{F}$ and  $\mathcal{B}$ respectively (see Fig.~\ref{fig: tripartizione grafo}): $V=\mathcal{I}\cup\mathcal{F}\cup\mathcal{B}$. More precisely, the sub-set $\mathcal{I}$ contains all vertices belonging to the fiber-graph $\mathcal{G}_i=\left(\mathcal{I},\mathcal{L}_i\right)$, the sub-set $\mathcal{F}$ contains all vertices belonging to the fiber-graph $\mathcal{G}_f=\left(\mathcal{F},\mathcal{L}_f\right)$ and finally the sub-set $\mathcal{B}$ contains all vertices belonging to the fiber-graphs $\mathcal{G}_k=\left(\mathcal{K}_k,\mathcal{L}_k\right)$ with $k\neq\left(i,f\right)$, formally $\mathcal{B}=\bigcup_{k\neq i,f}\mathcal{K}_k$, where $\mathcal{L}_j$ is the adjacency matrix of the fiber graph $\mathcal{G}_j$.

This procedure allows us to write the sum in (\ref{eq: Tetali}) as:
\begin{eqnarray}
\sum_{z\in\mathcal{V}}g\left(z\right)\left(R_{f,z}-R_{i,z}\right)  & = & \sum_{z\in\mathcal{I}}g\left(z\right)\left(R_{f,z}-R_{i,z}\right)+\nonumber \\
														& + & \sum_{z\in\mathcal{F}}g\left(z\right)\left(R_{f,z}-R_{i,z}\right)+\nonumber \\
														& + & \sum_{z\in\mathcal{B}}g\left(z\right)\left(R_{f,z}-R_{i,z}\right).
														\label{eq: 2) Decomp. Sum}
\end{eqnarray}
namely we distinguish the case where $z$ is in the same fiber graph as $i$ ($z\in\mathcal{I}$), $z$ is in the same fiber graph as $f$ ($z\in\mathcal{F}$), and $z$ is in a fiber graph which neither $i$ nor $f$ belong to ($z\in\mathcal{B}$). In the following we will call root $r_{i}$ the vertex shared by the base graph $\mathcal{G}_{0}$ and the fiber graph $\mathcal{G}_{i}$. 

In general, when two vertices $i$ and $f$ belong to different fiber-graphs, we can write the effective resistance $R_{i,f}$ between these point as a sum of three resistances in series:
\begin{equation}
R_{i,f}=R_{i,r_{i}}+R_{r_{i},r_{f}}+R_{f,r_{f}}.\label{eq:2) Decomp. R}
\end{equation}

Now, recalling Eq. (\ref{eq: 2) Decomp. Sum}), we can notice that when $z\in\mathcal{B}$ we can use Eq. (\ref{eq:2) Decomp. R}), for both $R_{i,z}$ and $R_{z,f}$, while when $z\in\mathcal{I}$ or $z\in\mathcal{F}$, we can use Eq. (\ref{eq:2) Decomp. R}) only for $R_{i,z}$ or for $R_{i,z}$, respectively. 

In fact, suppose we consider a linear chain as a fiber graph $\mathcal{G}_i$: when $z$ and $i$ belong to the same fiber graph the effective resistance according to Eq. (\ref{eq: 2) Decomp. Sum}) would be $|y_i|+|y_z|$, being $y_i$ and $y_f$ the coordinates of $i$ and $f$ along the related theeth, while the correct estimte is $|y_i-y_f|$, which recovers the former only when the signs are different. 

Exploiting this remark, the three sums appearing in Eq. (\ref{eq:2) Decomp. R}) can be written as: 
\begin{eqnarray*}
\sum_{z\in B}g\left(z\right)\left(R_{f,z}-R_{i,z}\right) & = & \sum_{z\in B}g\left(z\right)\left\{ R_{r_{f},r_{z}}-R_{r_{i},r_{z}}\right. +\\
 &  & +\left. R_{f,r_{f}}-R_{i,r_{i}}\right\} ,\\
\sum_{z\in I}g\left(z\right)\left(R_{f,z}-R_{i,z}\right) & = & \sum_{z\in I}g\left(z\right)\left\{ R_{f,r_{f}}+R_{r_{f},r_{z}}\right.+\\
 &  & +\left.R\left(z,r_{z}\right)-R_{i,z}\right\} ,\\
\sum_{z\in F}g\left(z\right)\left(R_{f,z}-R_{i,z}\right) & = & \sum_{z\in F}g\left(z\right)\left\{ R_{f,z}-R_{i,r_{i}}\right.+\\
 &  & -\left.R_{r_{i},r_{z}}-R_{z,r_{z}}\right\} .
\end{eqnarray*}
Then, after some algebraic manipulations we find that
\begin{eqnarray}
H\left(i,f\right) & = & mR_{r_{i},r_{f}}+\frac{1}{2}\sum_{z\in \mathcal{V}} g\left(z\right)\left(R_{r_f,r_z}-R_{r_i,r_z}\right)+\label{eq: Hbr Parziale}\\
 &  & +2\left(m-m_{f}\right)R\left(f,r_{f}\right)+H_{i}\left(i,r_{i}\right)+H_{f}\left(r_{f},f\right),\nonumber 
\end{eqnarray}
where $m_{k}$ is the number of links in the sub-graph $\mathcal{G}_{k}$, and  $H_{k}\left(a,b\right)$ the mean time for a walker to go from $a$ to $b$, $\left(a,b\in\mathcal{G}_0\right)$, without ever exiting out from the fiber $\mathcal{G}_0$.

We have obtained a summation similar to that with which we began, but now resistances are considered between points of the base-graph. This allows us to factorize the difference $\left(R_{r_f,r_z}-R_{r_i,r_z}\right)$, by splitting the summation over $z\in V$ into the summation over every vertex of all fiber-graphs: $$\sum_{z\in\mathcal{V}}g\left(z\right)\left(R_{r_f,r_z}-R_{r_i,r_z}\right)=\sum_{k\in\mathcal{G}_{0}}\sum_{z\in\mathcal{G}_{k}}g\left(z\right)\left(R_{r_f,r_z}-R_{r_i,r_z}\right).$$ Previously we noticed that the difference of resistances depends only on the fiber graph, so we can factorize it: 

\begin{eqnarray}
&   &  \sum_{z\in \mathcal{V}}g\left(z\right)\left(R_{r_f,r_z}-R_{r_i,r_z}\right)=\nonumber \\
& = & \sum_{k\in\mathcal{G}_{0}}\left(R_{r_f,r_k}-R_{r_i,r_k}\right)\sum_{z\in\mathcal{G}_{k}}g\left(z\right)\nonumber \\
																			& = & \sum_{k\in\mathcal{G}_{0}}\left(R_{r_f,r_k}-R_{r_i,r_k}\right)2m_{k}. \nonumber 
\end{eqnarray}

Finally, merging the previous results we get
\begin{eqnarray}
H\left(i,f\right)	& = & H_i \left(i,r_i\right) +\sum_{k\in \mathcal{G}_0} m_k \left(R_{r_f,r_k}-R_{r_k,r_i}\right)+\nonumber \\
 					& + & H_f\left(r_f,f\right)+2\left(m-m_f\right) R\left(f,r_f\right) \nonumber \\
					& + & mR_{r_{i},r_{f}}. \label{eq: H ramificato} 					
\end{eqnarray}

Of course, this formula recovers the Tetali one (\ref{eq: Tetali}) when all the branched graphs $\mathcal{G}_i$ are composed only of a single vertex (the one that is shared with $\mathcal{G}_{0}$, i.e. the root $r_i$). The details of the fibers only enter in the summation term, and the only important things about these graphs are the position on the base graph and the number of links $m_{k}$, neither the topology nor the number of vertices of these fiber graphs affect $H\left(i,f\right)$.

\subsection{Particular Cases:}

There are special cases where the formula in Eq. (\ref{eq: H ramificato}) can be significantly simplified. In particular when the ratio between the coordination number $g(k)$ of any vertex $k$ belonging to the base graph and the number of links $m_{k}$ in the $k-$th subgraph is independent of $k$. This allows us to write $m_{k}=\gamma g(k)$, being $\gamma$ a constat value which is the same for every vertex of the base graph $\mathcal{G}_0$. Then, the summation in Eq. (\ref{eq: H ramificato}) becomes:

\begin{equation}
\sum_{k\in\mathcal{G}_{0}}m_k\left(R_{r_f,r_z}-R_{r_i,r_z}\right)=2\gamma\left[H_{0}\left(r_{i},r_{f}\right)-m_{0}R_{r_{f},r_{i}}\right], \label{eq: futuranullaa}
\end{equation}
and, using this result, Eq. (\ref{eq: H ramificato}) becomes:
\begin{eqnarray}
H\left(i,f\right) & = & H_{I}\left(i,r_{i}\right)+2\gamma H_{0}\left(r_{i},r_{f}\right)+H_{F}\left(r_{f},f\right)+\label{eq: per pettine chiuso}\\
 &  & +2\left(m-m_{f}\right)R\left(f,r_{f}\right)+\left(m-2\gamma m_{0}\right)R_{r_{i},r_{f}},\nonumber 
\end{eqnarray}
where $ H_{I}\left(a,b\right)$ and $ H_{F}\left(a,b\right)$ represent the mean time taken by a walker to first go from a node $a$ to a node $b$ (being $a,b\in\mathcal{I}$ and $a,b\in\mathcal{F}$ respectively) without ever eciting out of $\mathcal{I}$ and $\mathcal{F}$ respectively; analogously $ H_0\left(a,b\right)$ represent the mean time to first reach $b$ starting from $a$ (being $a,b\in\mathcal{G}_0$) and moving only on the base graph.
In this case the mean time to go from $i$ to $f$ can be expressed by a sum of three hitting times, and some constant value.

\begin{figure}
	\includegraphics[width=\linewidth]{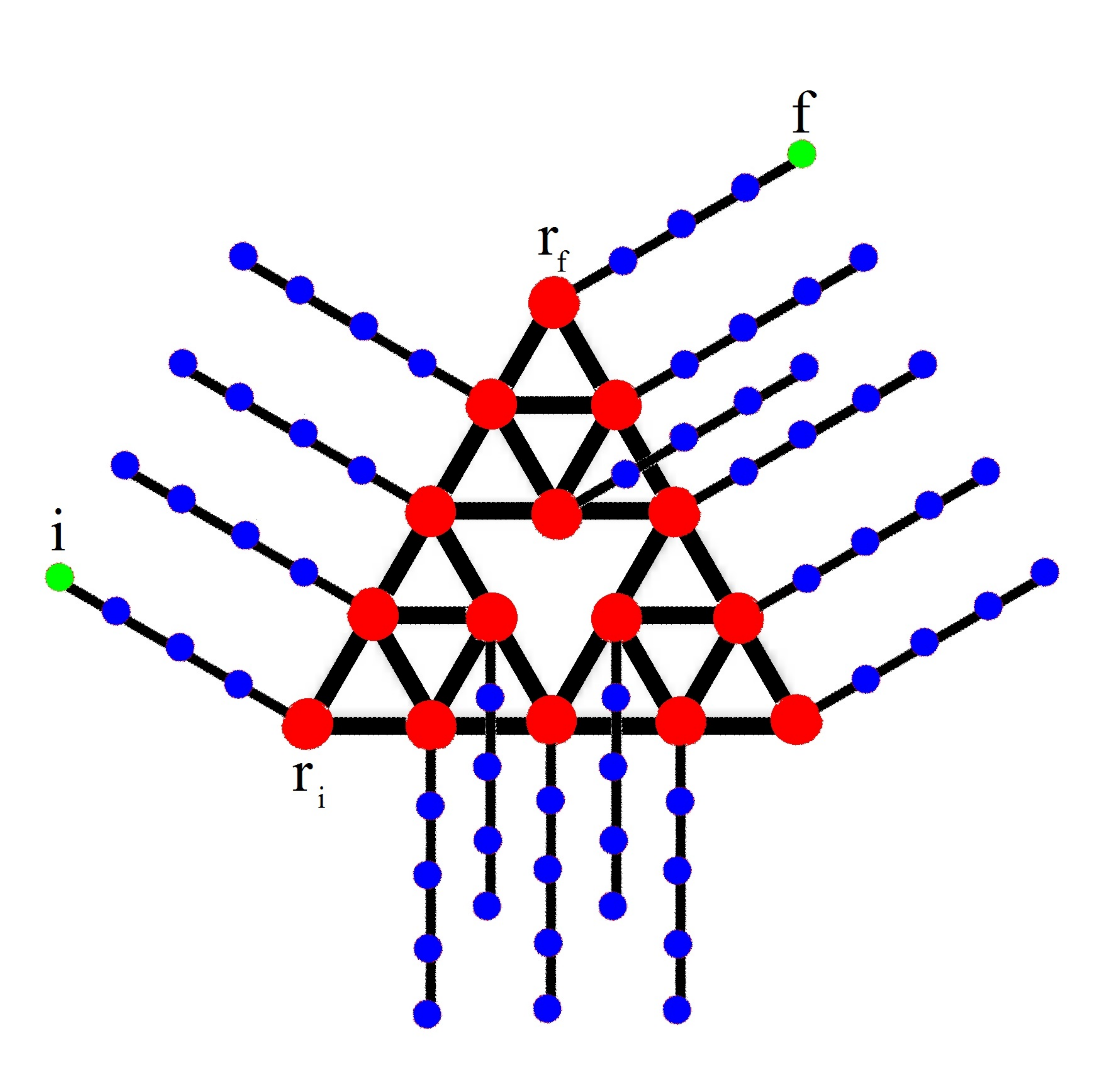}  
	\caption{Example of a Sierpinski gasket of generation 3 fibrated with linear chains with $5$ vertices.} \label{fig: Sierp}
\end{figure}

Another interesting case appears when the position of the root of the starting vertex $r_i$ and the root of the ending one $r_f$ displays a symmetry such that 
\begin{equation}
	H\left(r_i,r_f\right)=H\left(r_f,r_i\right) \label{Condicio}
\end{equation}
in this particular case Eq. (\ref{eq: H ramificato}) becomes
\begin{eqnarray}
H(i,f)& = & H_{i}\left(i,r_{i}\right)+H_{f}\left(r_{f},f\right)+ \nonumber\\
&+&2\left(m-m_{f}\right)R\left(f,r_{f}\right)+mR_{r_{i},r_{f}}. \label{eq: Piusemplice}
\end{eqnarray}
This formula allows us to calculate the hitting times for any "symmetric structure" (according to Eq. (\ref{Condicio})), like a Sierpinski fibrated with linear chains, (see Fig \ref{fig: Sierp}).

Let us consider a Sierpinski gasket of generation $n$ and as fibers linear chains with $L$ vertices. The total number of vertices in the base-graph is \cite{Sierpinsky} $|V|^{(n)}_0= \frac{3 \left(3^n +1 \right)}{2}$ and the total number of links is \cite{Sierpinsky} $m^{(n)}_0= 3^{n+1}$. For instance let us try to evaluate the mean time to first reach the node $f$ belonging to a tooth stemming from a corner, starting from a node $i$ belonging to a tooth stemming from another corner (see Fig \ref{fig: Sierp}). Neri et al. in \cite{Sierpinsky2} showed that $R_{r_i,r_f}=\frac{2\cdot 5^n}{3^{n+1}}$. Using Eq. (\ref{eq: Piusemplice}) it is straightforward to show that $H\left(i,f\right)$ reads as

\begin{equation*}
H_{\textrm{Max}} = 3L^2+\left(1+\frac{L}{2}\right)\left(3^{n+1}+2\cdot 5^{n}\right)+\left(\frac{5}{3}\right)^n L .
\end{equation*}

\subsection{Bidimensional Combs} \label{sub:UNICUM}

Bidimensional combs are branched structures where the base-graph $\mathcal{G}_{0}$ is a monodimensional graph, like a linear chain or a ring, usually called backbone, and fiber graphs are linear chains usually called teeth, see Fig. \ref{fig: Pettini}. Here, we consider two different cases according to the the boundary conditions applied to the backbone: we call ``Bidimensional Open Combs'' those combs whose backbone is a linear chain, and ``Bidimensional Closed Combs'' those whose backbone is a ring; teeth are always taken as open (i.e. reflecting at boundaries). 

As shown in Fig. \ref{fig: Pettini}, the size of the backbone is $2L+1$, and each linear chain departing from the backbone counts $\alpha L$ vertices, in such a way that $\alpha$ measures the degree of inhomogeneity along the two directions (e.g., when $\alpha=1$ the comb is square).

Using Eq. (\ref{eq: H ramificato}) we are now able to calculate the value of $H\left(i,f\right)$ for those graphs.

In the following, exploiting the fact that combs are embedded in $2-$dimensional lattices, the position of the arbitrary point $k$ will be denoted as $\left(x_k,y_k\right)$ where $x_k$ indicates the projection of k on the backbone, and $y_k$ its length along the related tooth.
Let us start with open combs for which we apply Eq. (\ref{eq: H ramificato}), with some algebra we get \footnote{What we really find from Eq. (\ref{eq: H ramificato}) is $H\left(i,f\right)-2L\left(1+\alpha+2\alpha L\right)\left[-1+|y_{i}|+|y_{f}|-|y_{i}-y_{f}|\right]\delta_{x_{i},x_{f}}$. This difference is due to the hypotesis that $x_i\neq x_f$, namely the starting point and the final point belong to different fiber graphs. If we use Eq. (\ref{eq: per pettine chiuso}) instead of Eq. (\ref{eq: H ramificato}), we introduce an additional error of $\left(x_{i}-x_{f}\right)\left(x_{i}+x_{f}\right)$ due the different value of $g\left(x,0\right)$ between $x=\pm L$, where $g\left(\pm L,0\right)=3$, and $x\neq\pm L$, where $g\left(x,0\right)=4$.}:
\begin{eqnarray}
\nonumber 
H\left(i,f\right) & = & mR_{i,f}+\left(y_{f}^{2}-y_{i}^{2}\right)+\left(2\alpha L+1\right)\left(x_{f}^{2}-x_{i}^{2}\right)+\\
\label{eq: Comb 2D Open}
 & +& \left(|y_{2}|-|y_{1}|\right)\left(L^{2}4\alpha+2L\right),
\end{eqnarray}
where the resistance between two generic points $a=\left(x_a,y_a\right)$ and $b=\left(x_b,y_b\right)$ for open combs is $$R_{a,b}=\delta_{x_a,x_b}|y_a-y_b|+\left(1-\delta_{x_a,x_b}\right)\left(|x_a-x_b|+|y_a|+|y_b|\right).$$

From this equation we can extract important information, for example the mean time to cross the whole backbone and the mean time to ``climb'' a tooth. These quantities are, respectively,
\[
H\left({ -L,0}\rightarrow {L,0} \right)=L^{3}\left(8\alpha\right)+L^{2}\left(4+4\alpha\right),
\]
and
\[
H\left(\left\{ 0,0\right\}\rightarrow\left\{ 0,\alpha L\right\} \right)=L^{3}\left(8\alpha^{2}\right)+L^{2}\left(4\alpha+3\alpha^{2}\right).
\]

Using Eq. (\ref{eq: H ramificato}) we can also calculate the Hitting Time for closed combs, $H_{2}^{\circlearrowleft}\left(i,f\right)$, where we used the apex $^\circlearrowleft$ to emphasize that this quantity is reffered to a closed comb
\footnote{Once again what we really find using Eq. (\ref{eq: H ramificato}) is $H_{2}^{\circlearrowleft}\left(i,f\right)-\left(1+2L\right)\left(1+2\alpha L\right)\left[|y_{i}|+|y_{f}|-|y_{i}-y_{f}|\right]\delta_{x_{i},x_{f}}$. This difference is due to the hypotesis that $x_i\neq x_f$. This time if we use Eq. (\ref{eq: per pettine chiuso}) instead of Eq. (\ref{eq: H ramificato}), we do not introduce any additional error.}:
\begin{equation}
H_{2}^{\circlearrowleft}\left(i,f\right)=mR_{if}^\circlearrowleft+\left(|y_{f}|-|y_{i}|\right)\left[4\alpha L^{2}+2L+1\right]+\left(y_{f}^{2}-y_{i}^{2}\right),\label{eq: Closed Comb 2d}
\end{equation}
where the resistance between two generic points $a=\left(x_a,y_a\right)$ and $b=\left(x_b,y_b\right)$ for close combs is 

\begin{eqnarray*}
R_{a,b}^{\circlearrowleft} & = &\delta_{x_a,x_b}|y_a-y_b|+\left(1-\delta_{x_a,x_b}\right)\times\\
		&   & \times\left(|x_a-x_b|+|y_a|+|y_b|-\frac{\left(x_a-x_b\right)^2}{2L+1}\right).
\end{eqnarray*}From Eq. (\ref{eq: Closed Comb 2d}) we can extract the time needed to cross half backbone as
\[
H_{2}^{\circlearrowleft}\left(\left\{ 0,0\right\} \rightarrow\left\{ L,0\right\} \right)=L^{3}\left(6\alpha\right)+L^{2}\left(3+2\alpha\right)+L,
\]
and the time to climb a thoot as
\[
H_{2}^{\circlearrowleft}\left(\left\{ 0,0\right\} \rightarrow\left\{ 0,\alpha L\right\} \right)=L^{3}\left(8\alpha^{2}\right)+L^{2}\left(4\alpha+3\alpha^{2}\right)+L\left(2\alpha\right).
\]
We can notice that the leading order of the mean time needed to cross the backbone is proportional to $\alpha$ (consistent with results in \cite{Redner1}) and the leading order of the mean time to climb a tooth is proportional to $\alpha^{2}$.

\subsection{$d$-dimensional Open Combs}

\begin{figure}
	\includegraphics[width=\linewidth]{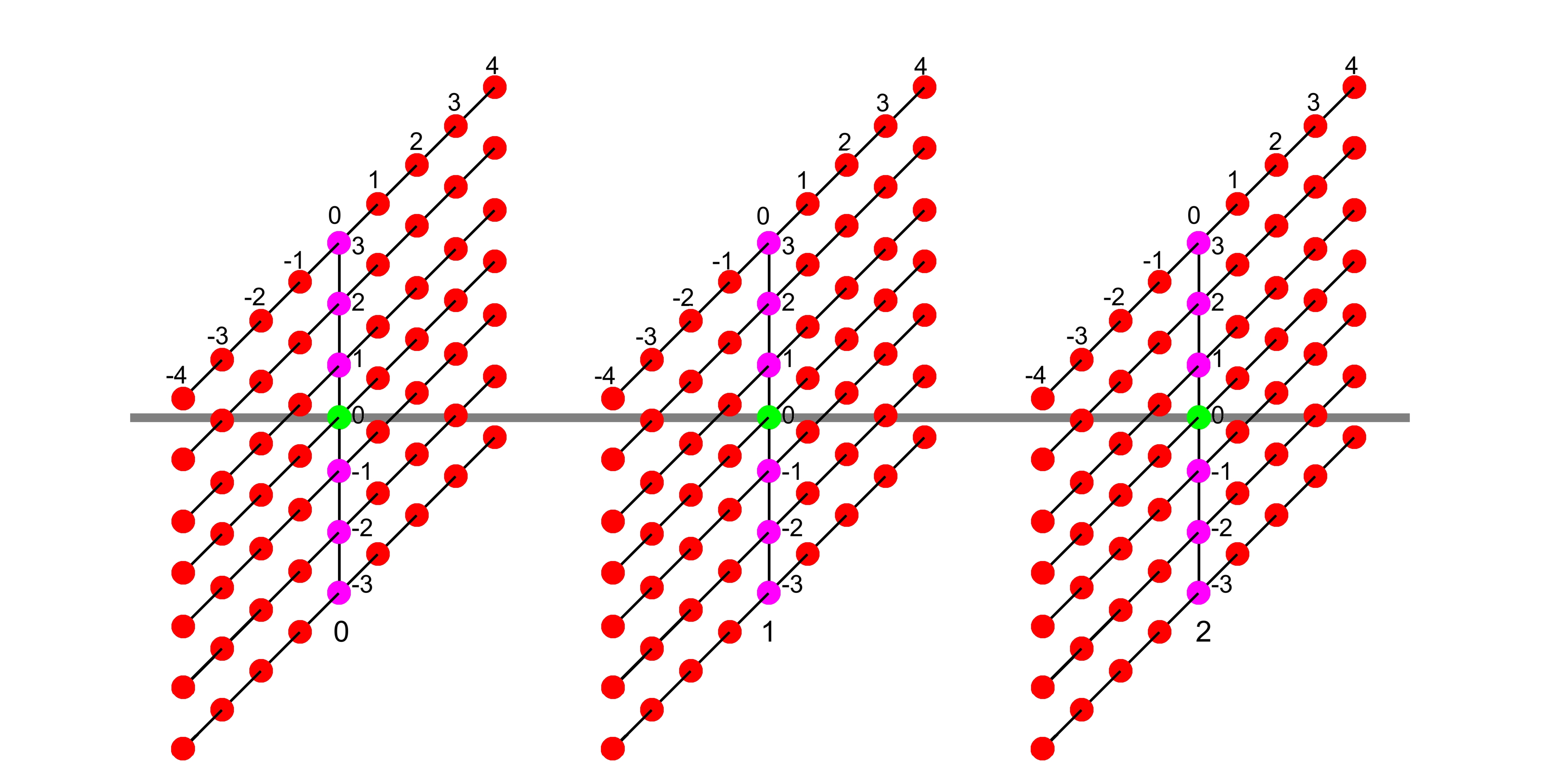}  
	\caption{Example of a 3-dimensional comb. Vertices which belong to the first generation are green, those which belong to the second are lilac, and those which belong to the third are red.} \label{fig: TRIDI}
\end{figure}

We define $d$-dimensional open combs recursively (see also Fig \ref{fig: TRIDI}):
	\begin{itemize}
	\item a $1-$dimensional open comb is a linear chain;
	\item a $2-$dimensional open comb is a branched graph $\mathcal{G}$ whose base graph is a linear chain and whose fiber graphs are linear chain;
	\item \ldots
	\item a $d-$dimensional open comb is a branched graph $\mathcal{G}$ whose base graph is a $(d-1)-$dimensional combs and whose fibers are linear chains.
	\end{itemize}
As shown in Sec. \ref{sub:UNICUM}, a finite $2$-dimensional comb can be defined by fixing two parameters: the length of the backbone $L$ and the ratio $\alpha$ between the length of a tooth and the length of the backbone, in such a way that the thermodynamic limit $L \rightarrow \infty$ is well defined. Now, to fix a $d-$dimensional comb we introduce $d$ parameters: $\left(L,\alpha_2,\ldots,\alpha_d \right)$, being $\alpha_i$ the ratio between the length of the tooth in the $i$-th direction (i.e. added at the $i-$th iteration) and the length of the backbone: $\alpha_i=\frac{L_i}{L}$.

We label every vertices with $d$ coordinates and call $i=\left(x_{1},\ldots,x_{d}\right)$ and $f=\left(y_{1},\ldots,y_{d}\right)$, where the first coordinate labels the vertices on the base graph and it takes value from $-L$ to $L$, the second one goes from $-\alpha_{2}L$ to $\alpha_{2}L$ and labels the base of the fiber graphs, the third one goes from $-\alpha_{3}L$ to $\alpha_{3}L$ and so on.

We define $H^{\left(d \right)}\left(i,f \right)$, the Hitting Time from $i$ to $f$ in a $d-$dimensional comb and we will use Eq. (\ref{eq: H ramificato}) to calculate its leading order. Using Tetali's equation it is straightforward to observe that $H^{\left(d \right)}\left(i,f \right) \sim \mathcal{O}\left(L^{d+1}\right)$. In fact, the first term of the right hand side in Eq. (\ref{eq: Tetali}) is $m_{d}R_{i,f}$ and one can see that in the $d-$dimensional comb  the number of links $m_{d}$ is
\begin{equation}
m_{d}\sim\left(L^{d}2^{d}\prod_{k=2}^{d}\alpha_{k}\right),\label{eq: M_n}
\end{equation}
while the maximum value of $R$ is
\[
R=2L\left(1+\sum_{k=2}^{d}\alpha_{k}\right)+1,
\]
thus, the leading order of this first term is $L^{d+1}$. Also the second term in the right hand side of the Eq. (\ref{eq: Tetali}), that is $\frac{1}{2}\sum_{z\in V}g\left(z\right)\left(R_{f,k}-R_{k,i}\right)$, where $g\left(z\right)\leq2d=\mathcal{O}\left(L^{0}\right)$, $V=\mathcal{O}\left(L^{d}\right)$ and $R=\mathcal{O}\left(L\right)$, is order $L^{d+1}$.

Using this observation we can simplify Eq. (\ref{eq: H ramificato}), in fact $H_{i}$, $H_{f}$ and $m_{d}R\left(f,r_{f}\right)$ are $\mathcal{O}\left(L^{d}\right)$. So the leading order of Eq. (\ref{eq: H ramificato}) is given only by
\begin{eqnarray}
H\left(i,f\right) 	& \sim & \frac{1}{2}\sum_{k=0}^{L}\left\{ |k-y_{1}|-|k-x_{1}|\right\} m_{k}+\nonumber \\
					&  +   & 2m_{d}R\left(f,r_{f}\right)+m_{d}R_{r_{i},r_{f}}.
\end{eqnarray}
We have just pointed out that $m_{d}\sim\left(L^{d}2^{d-1}\prod_{k=2}^{d}\alpha_{k}\right)$, where $m_{k}$ is the number of links in the fiber-graphs starting from the node $k$, and its leading value is $m_{k}\sim\left(L^{d-1}2^{d-1}\prod_{k=2}^{d}\alpha_{k}\right)$,
so
\begin{eqnarray*}
H\left(i,f\right)^{\left(d\right)} & \sim & \left(L^{d-1}2^{d-2}\prod_{k=2}^{d}\alpha_{k}\right)\left(y_1^2-x_1^2\right) +\\
 &  & +\left(L^{d}2^{d}\prod_{k=2}^{d}\alpha_{k}\right)\left[|x_{1}-y_{1}|+2\sum_{j=2}^{d}|y_{j}|\right].
\end{eqnarray*}

By introducing the normalized coordinates $\xi_{i}=\frac{y_{i}}{\alpha_{i}L}$, and $\eta_{i}=\frac{x_{i}}{\alpha_{i}L}$ (posing conventionally $\alpha_1=1$), we can express the leading value of $H^{\left(d\right)}\left(i,f\right)$:
\begin{eqnarray}
H\left(\overrightarrow{\eta},\overrightarrow{\xi}\right) & \sim 	& L^{d+1}2^{d-2}\prod_{i=2}^{d}\alpha_{i}\times \label{eq: ASIM}\\
 														 & \times	& \left[\left(\xi_{1}^{2}-\eta_{1}^{2}\right)+4|\xi_{1}-\eta_{1}|+8\sum_{k=2}^{d}\alpha_{k}\xi_{k}\right].\nonumber
\end{eqnarray}
We can use Eq. (\ref{eq: ASIM}) to calculate the mean time to cross the backbone going from $\eta=\left(-1,\vec{0}\right)$ to $\xi=\left(+1,\vec{0}\right)$, as
\[
H^{\left(d\right)}\left(\left\{{-1,\vec{0}}\right \}\rightarrow \left\{1,\vec{0}\right\}\right)\sim L^{d+1}2^{d+1}\prod_{i=2}^{d}\alpha_{i},
\]
and its maximum value going from $\eta=\left(-1,\vec{1}\right)$ to $\xi=\left(1,\vec{1}\right)$:
\[
H^{\left(d\right)}\left(\left \{{-1,\vec{1}}\right \}\rightarrow \left\{+1,\vec{1}\right\}\right)\sim L^{d+1}2^{d+1}\prod_{i=2}^{d}\alpha_{i}\left[1+\sum_{k=2}^{d}\alpha_{k}\right].
\]

\section{Trapping Times}

In this section, we will analyze some average quantities, in particular we introduce the mean first-passage time defined by $\mbox{MFPT}_{f}=\mathbb{E}_{i}\left[H\left(i,f\right)\right]$ and the Global Mean First Passage Time, defined by $\mbox{GMFPT}=\mathbb{E}_{i}\left[\mathbb{E}_{f}\left[H\left(i,f\right)\right]\right]$. The first quantity rapresents the mean time to reach a fixed reaction node placed in $f$, starting from a random one. The second one is the mean time to reach a random vertex starting from another random one.

\subsection{MFPT$_{f}$}

The $\mbox{MFPT}_{f}$ was introduced by Montroll for regular graphs \cite{Montroll} to describe the excitation transfer on a photosintetic complex. Later this quantities was extended for irregular graph, see \cite{Benichou1,Redner1,MFPT1,MFPT2,MFPT3,MFPT4}. Once we know every value of $H\left(i,f\right)$ we can calculate exactly this quantity using MFPT$_{f}=\frac{1}{V}\sum_{i\in\mathcal{V}}H\left(i,f\right)$, where $V$ is the volume of the graph $\mathcal{V}$. 

\subsubsection{Bidimensional Open Combs \label{sec: Open}}

Knowing the values of $H\left(i,f\right)$ for every couple of nodes $\left(i,f\right)$ (see Eq.\ref{eq: Comb 2D Open}), it is straightforward to calculate the value of $\mbox{MFPT}_{f}=\frac{1}{V}\sum_{i\in\mathcal{V}}H\left(i,f\right)$, where $V$ is the volume of the graph $\mathcal{V}$. 
To achieve this results it is convenient to split Eq. (\ref{eq: Comb 2D Open}) in three parts:
\begin{eqnarray*}
H\left(i,f\right)&=&\left\{ mR_{if}\right\} +\left\{ \left(2\alpha L+1\right)\left(x_{f}^{2}-x_{i}^{2}\right)\right\} +\\&&+\left\{ \left(y_{f}^{2}-y_{i}^{2}\right)+\left(|y_{f}|-|y_{i}|\right)\left(L^{2}4\alpha+2L\right)\right\} \\&=&\left(A\right)+\left(B\right)+\left(C\right),
\end{eqnarray*} and calculate these separately. The first contribution, $\mathbb{E}_i\left[A\right]$, can be written as

\begin{eqnarray*}
	\mathbb{E}_{i}\left[A\right]  	& = &  	\sum_{x_i=-L}^{L}\sum_{y_i=-\alpha L}^L m\left[\delta_{x_i,xf}|y_i-y_f|\right.+\\
									&  & +	\left.\left(1-\delta_{x_i,xf}\right)\left(|x_i-x_f|+|y_i|+|y_f|\right)\right],
\end{eqnarray*}
and after the summation on $x_i$ it becomes: 

\begin{eqnarray*}
\mathbb{E}_{i}\left[A\right]  	& = &	\frac{m}{V}\sum_{y_i=-\alpha L}^{\alpha L}\left[|y_i-y_f|+\right. \\
								&   & \left[+2L\left(|y_i|+|y_f|+x_f^2+L+L^2\right)\right],
\end{eqnarray*}
and finally
\begin{eqnarray*}
\mathbb{E}_{i}\left[A\right] 	& = &	y_f ^{2} \frac{m}{V}+|y_f| \left[ \frac{m}{V}\left(2\alpha L+1\right)2L\right]+\\
								&   & +	x_f^{2}\left[\frac{m}{V}\left(2\alpha L +1\right)\right]+\frac{m}{V}\left[\alpha L\left(\alpha L+1\right)+\right. \\
								&  & \left. +L\left(L+1\right)\left(2\alpha L+1\right) +2\alpha L^2\left(\alpha L+1\right)\right].
\end{eqnarray*}
Performing similar calculation the second contribution, $\mathbb{E}_{i}\left[B\right]$, can be written as
\begin{eqnarray*}
	\mathbb{E}_{i}\left[B\right] & = & \sum_{x_i,y_i} \left[ \left(2\alpha L+1\right)\left(x_{f}^{2}-x_{i}^{2}\right)\right]\\
								 &  & = \left(2\alpha L+1\right)x_f^2-\frac{1}{3}L\left(L+1\right)\left(2\alpha L+1\right); 
\end{eqnarray*}
and the third contribution, $\mathbb{E}_{i} \left[C\right]$, becomes:
\begin{eqnarray*}
\mathbb{E}_{i} \left[C\right]=|y_f|2L\left(2\alpha L+1\right)-y_f^2-\alpha L \left( \frac{1}{3}+2L\right).\\ 
\end{eqnarray*}
Merging these three contributions we find the exact value of $\mbox{MFPT}_f$:

\begin{eqnarray}
 &  & \mbox{MFPT}_{\left(x_{f},y_{f}\right)}  =  \label{eq: 3) MFPT open}\\
 &  & \left(1+\frac{m}{V}\right)\left[y_{f}^{2}+x_{f}^{2}\left(2\alpha L+1\right)\right.\left.|y_{f}|\left(4\alpha L^{2}+2L\right)\right]+\nonumber \\
 &  & +\left.|y_{f}|\left(4\alpha L^{2}+2L\right)\right]+\frac{1}{3\left(1+2L\right)\left(1+2\alpha L\right)}\times\nonumber \\
 &  & \times\left[16\alpha^{2}L^{5}+L^{4}\left(16\alpha+24\alpha^{2}+8\alpha^{3}\right)\right.+\nonumber \\
 &  & \left.-L\left(1+\alpha\right)+L^{2}3\left(1+\alpha\right)L^{3}\left(4+18\alpha+14\alpha^{2}+4\alpha^{3}\right)\right].\nonumber 
\end{eqnarray}

\begin{figure}[htbp]
	\centering
	\includegraphics[width=\linewidth]{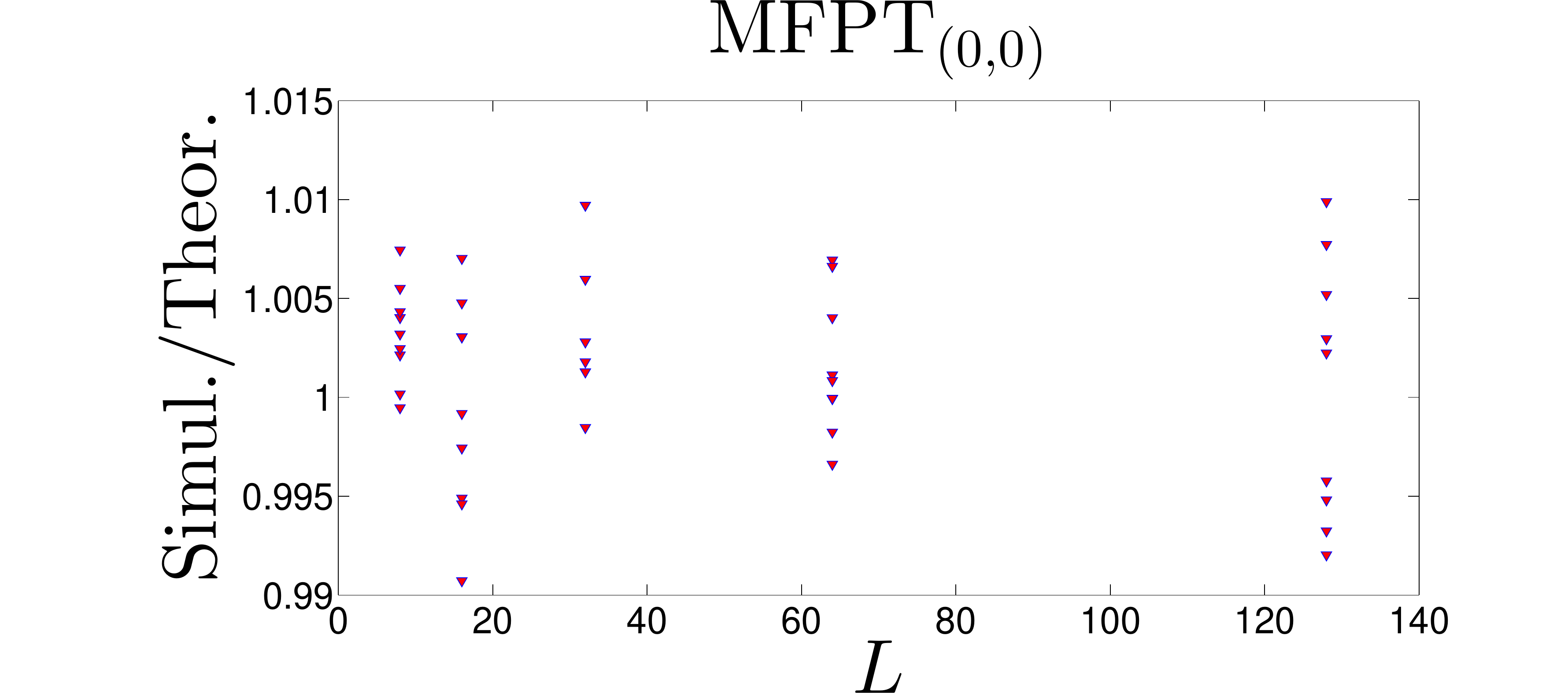} 
	\includegraphics[width=\linewidth]{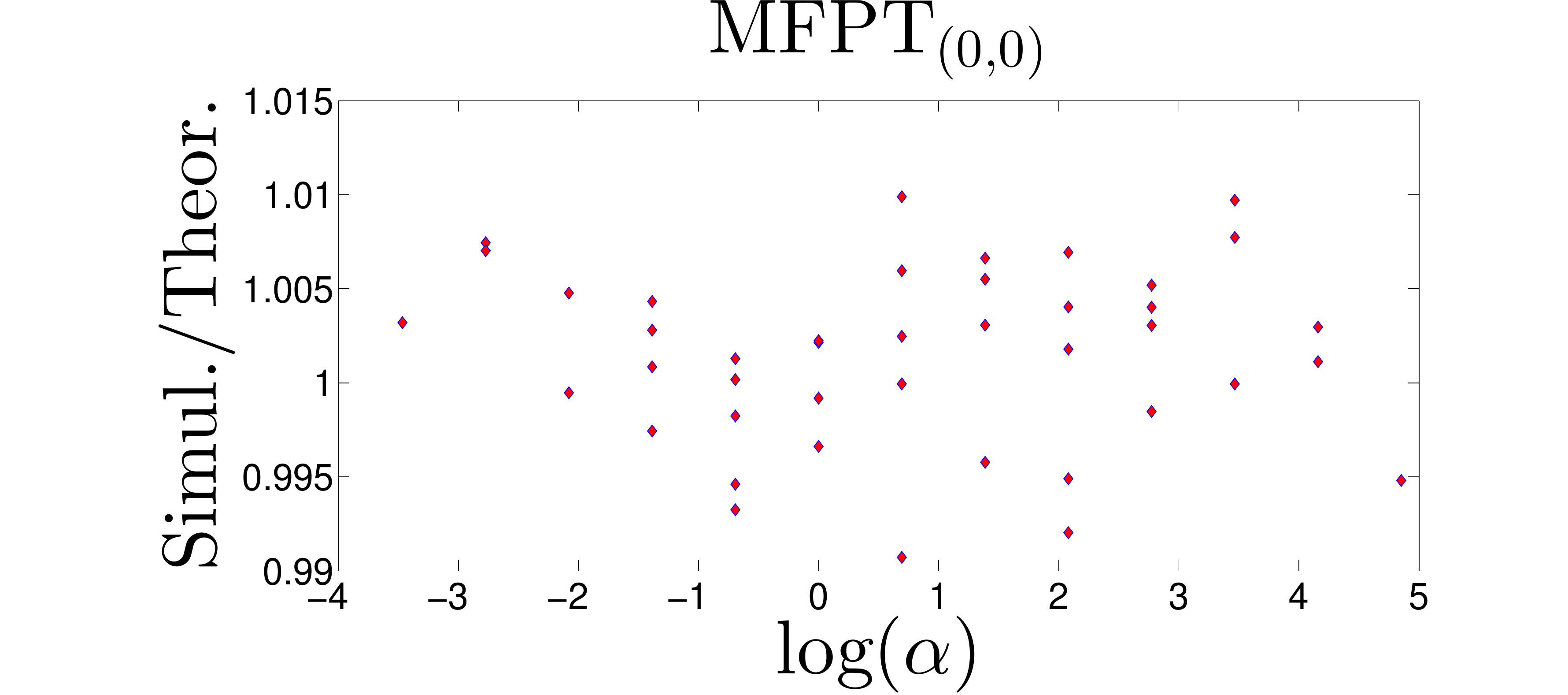} 
	\caption{\label{fig: MTT Dorso Open} Numerical check of the theoretical predictions of Eq. (\ref{eq: 3) MFPT open}) for $\left(x_{f},y_{f}\right)=\left(0,0\right)$. In both figures the results of the simulations are divided by the theoretical value of $\mbox{MFPT}_{\left(0,0\right)}$.}
\end{figure}

\begin{figure}[htbp]
\includegraphics[width=\linewidth]{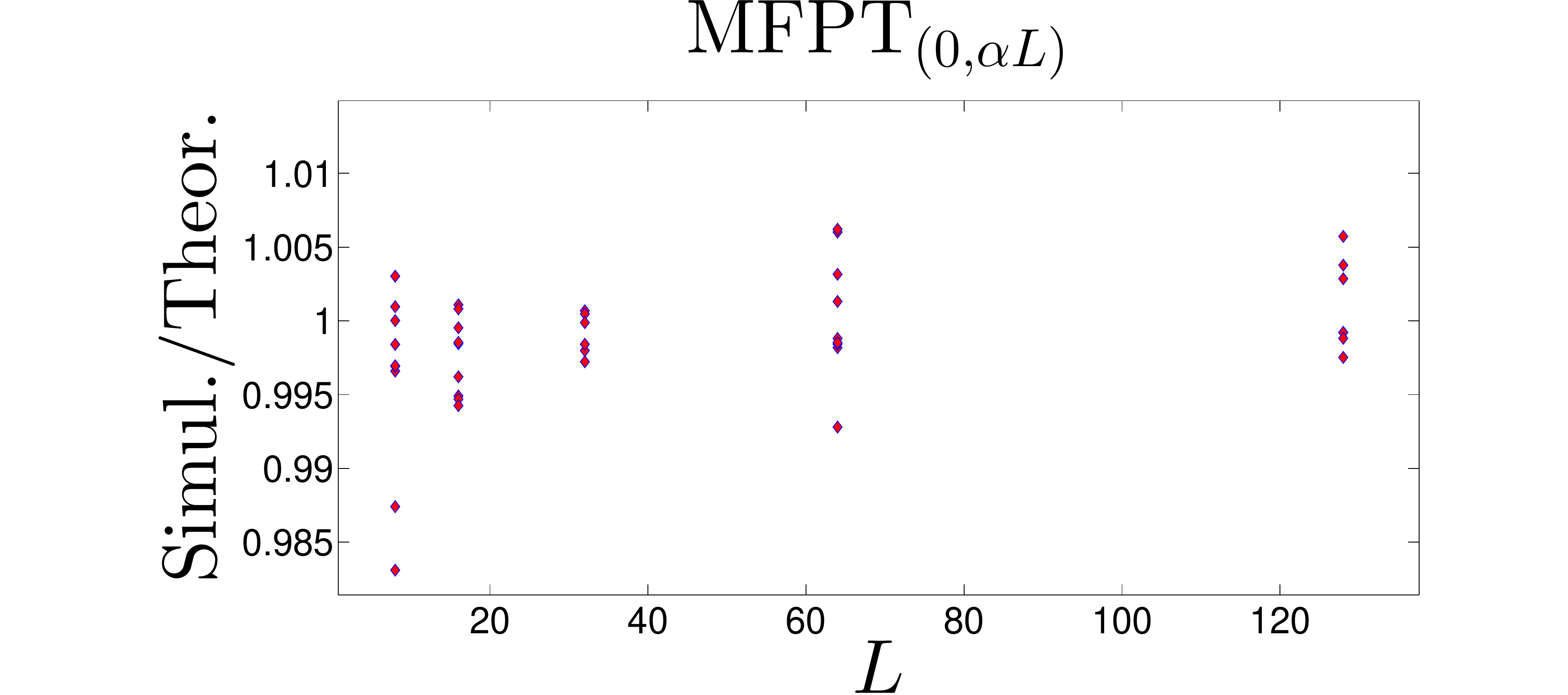}
\includegraphics[width=\linewidth]{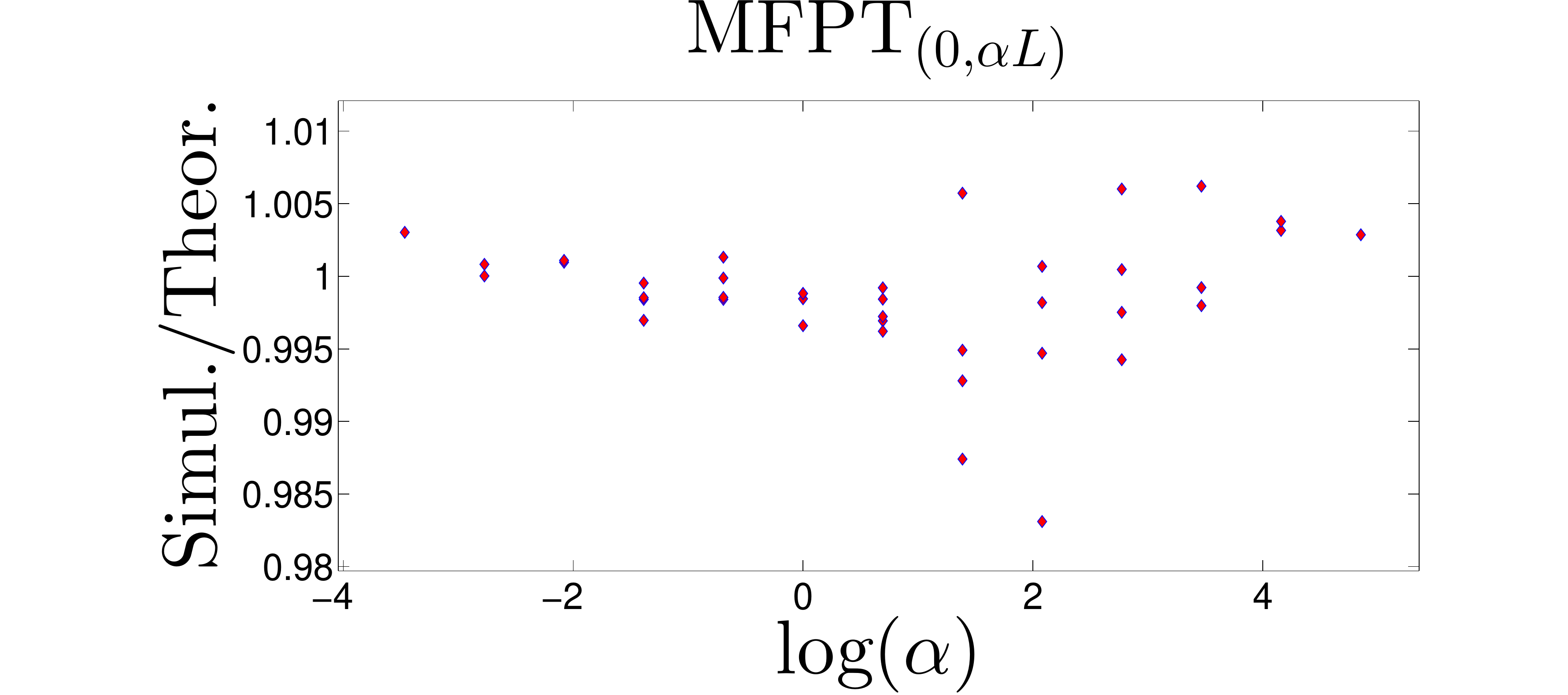}\protect\caption{\label{fig: MTT dente Open} Numerical check of the theoretical predictions of Eq. (\ref{eq: 3) MFPT open}) for $\left(x_{f},y_{f}\right)=\left(0,\alpha L\right)$. In both figures the results of the simulations are divided by the theoretical value of $\mbox{MFPT}_{\left(0,\alpha L\right)}$.}
\end{figure}

\begin{figure}[htbp]
\includegraphics[width=\linewidth]{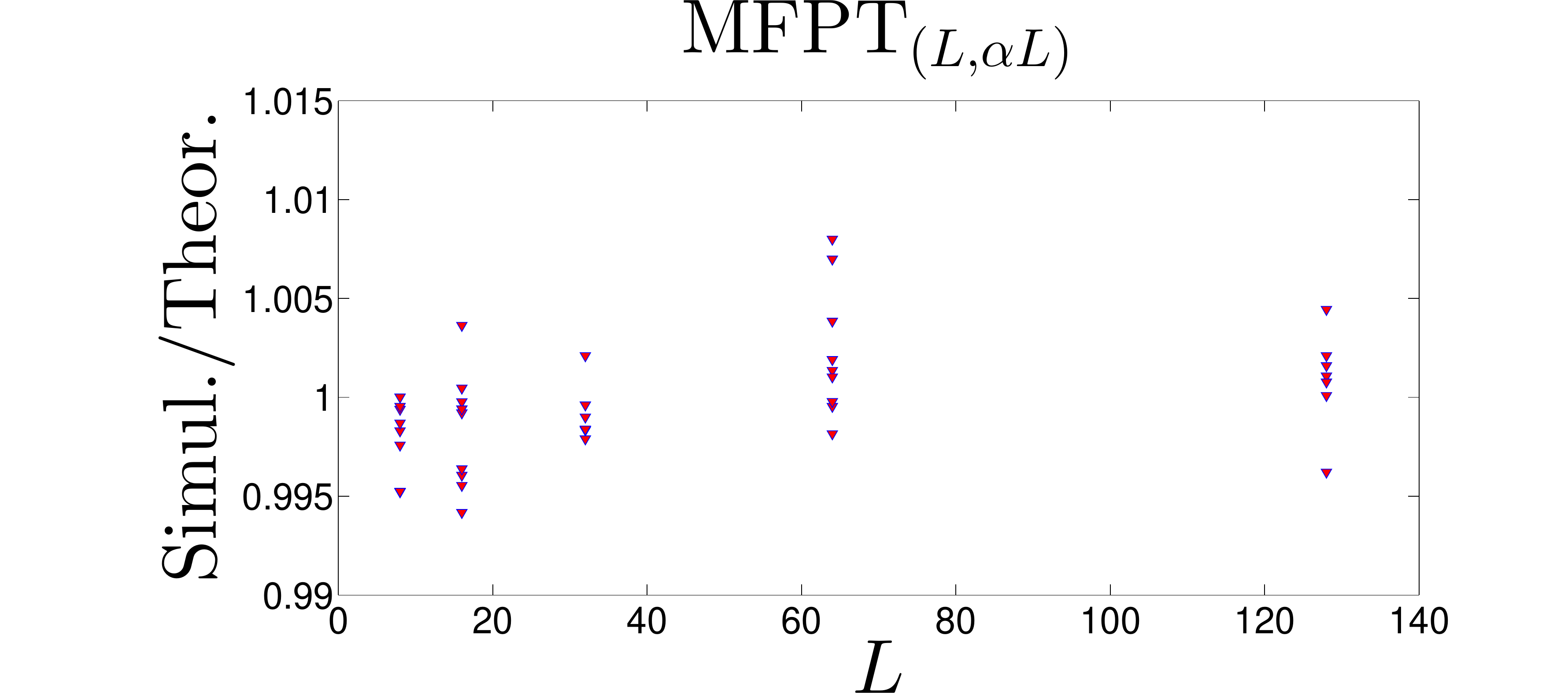}
\includegraphics[width=\linewidth]{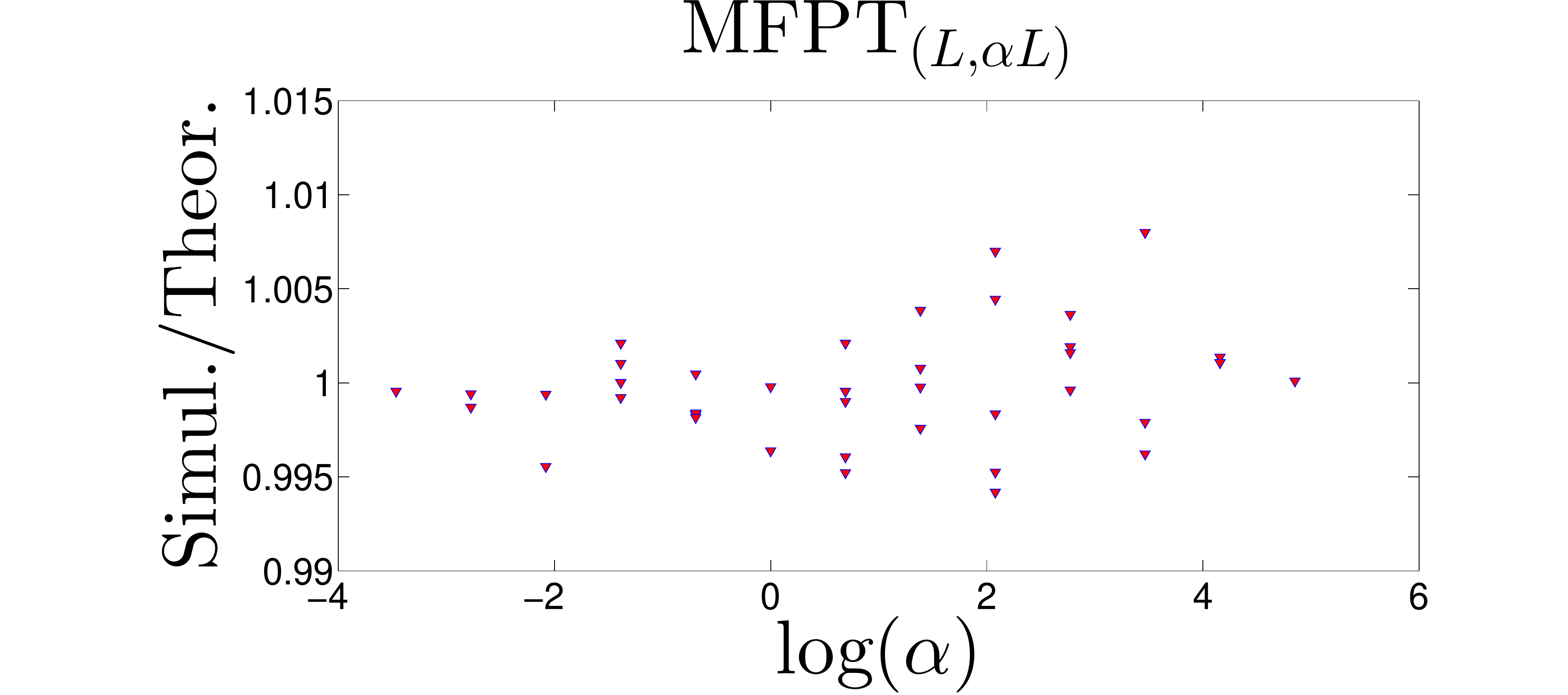}\protect\caption{\label{fig: MTT dente Open-MAX} Numerical check of the theoretical predictions of Eq. (\ref{eq: 3) MFPT open}) for $\left(x_{f},y_{f}\right)=\left(L,\alpha L\right)$. In both figures the results of the simulations are divided by the theoretical value of $\mbox{MFPT}_{\left(L,\alpha L\right)}$.}
\end{figure}

The position of the trap affects only the contribute in the first square braket. Therefore, $\mbox{MFPT}_{\left(x_{f},y_{f}\right)}$ can be written as the sum of two terms, the former depends on the final vertex, and the latter depends only on $L$ and $\alpha$ and is exactly the value of $\mbox{MFPT}_{\left(0,0\right)}$:
\begin{equation}
\mbox{MFPT}_{\left(x_{f},y_{f}\right)}\left(L,\alpha\right)=\phi\left(x_{f},y_{f}\right)+\mbox{MFPT}_{\left(0,0\right)}. \label{eq: MFPT_div}
\end{equation}
We have numerically tested this result for three positions of the trap $\left(x_{f},y_{f}\right)$ and for several values of $L$ and $\alpha$, as shown in Figs. (\ref{fig: MTT Dorso Open}, \ref{fig: MTT dente Open} and \ref{fig: MTT dente Open-MAX}). In each of those figures the value of $\left(x_{f},y_{f}\right)$ is fixed and we changed the value of $L$ and $\alpha$. The values of $\left(x_{f},y_{f}\right)$ are respectively $\left(0,0\right)$,$\left(L,0\right)$ and $\left(L,\alpha L\right)$, $L$ takes values from $8$ and $256$, and $\alpha$ goes from $1/32$ to $128$.

The asymptotic behaviour of $\mbox{MFPT}_{f}$ is
\begin{eqnarray} 
\mbox{MFPT}_{f}\left(L,\alpha\right) 	& \sim  & L^{3}\left\{ 4\alpha\left[\frac{1}{3}+\left(\frac{x}{L}\right)^2\right]+8\alpha \left| \frac{y}{L}\right|\right\} \label{eq: 3) AS MFPT} \\
										& \sim  & \frac{4}{3}\alpha L^3, \nonumber
\end{eqnarray}
where the last relation holds as long as both $x$ and $y$ are finite.
The dependence on $L^3$ is consistent with previous asymptotic results \cite{FCP4,Redner1}.

\subsubsection{Bidimensional Close Combs}

Analogous argoments can be applied to calculate the exact value of the Mean First Passage Time for close combs, and we call this quantity $\mbox{MFPT}_f^\circlearrowleft$. We skip lengthy passages and provide straightforwardly the final result:
\begin{eqnarray}
\mbox{MFPT}_{\left(x_{f},y_{f}\right)}^{\circlearrowleft} & = & \left[|y_{f}|\left(1+4L+8L^{2}\alpha\right)+2y_{f}^{2}\right]+\frac{1}{3+6\alpha L}\times\nonumber \\
 &  & \times\left[L\left(2-\alpha\right)+L^{2}\left(2+8\alpha+3\alpha^{2}\right)+\right.\nonumber \\
 &  & +\left.L^{3}4\alpha\left(2+2\alpha+\alpha^{2}\right)+L^{4}\left(8\alpha^{2}\right)\right].\label{eq: 3) MFPT close}
\end{eqnarray}
Of course, $\mbox{MFPT}^\circlearrowleft_f$ does not depend on $x_{f},$ due to the periodic condition on the $x$ axis. 
Moreover, as done in Eq. (\ref{eq: MFPT_div}) we can distinuish two contribution highlighting the dependance of the trap position (height) as

\begin{equation*}
\mbox{MFPT}_{\left(x_{f},y_{f}\right)}^\circlearrowleft \left(L,\alpha\right)=\phi^\circlearrowleft \left(y_{f}\right)+\mbox{MFPT}_{\left(0\right)}.
\end{equation*}

\begin{figure}[htbp]
\includegraphics[width=\linewidth]{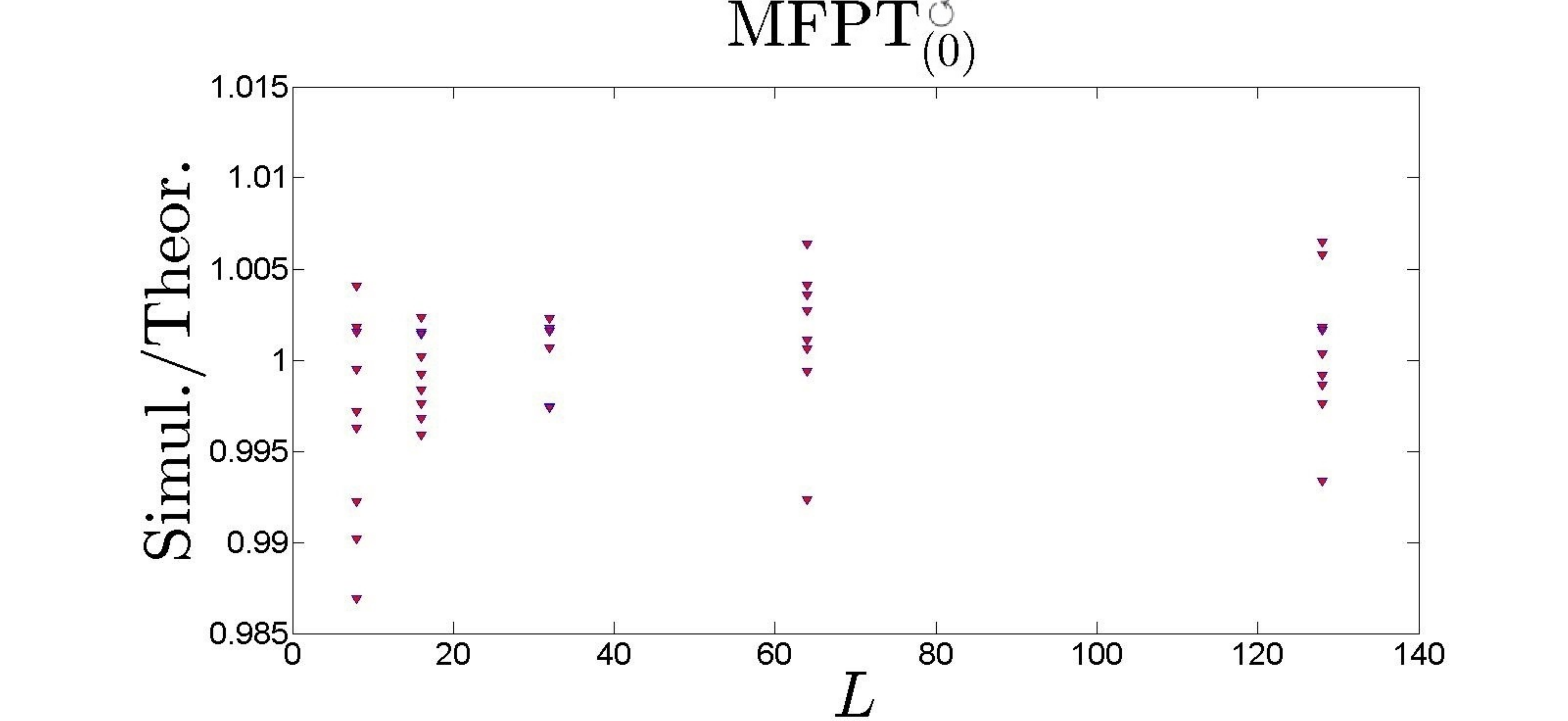}
\includegraphics[width=\linewidth]{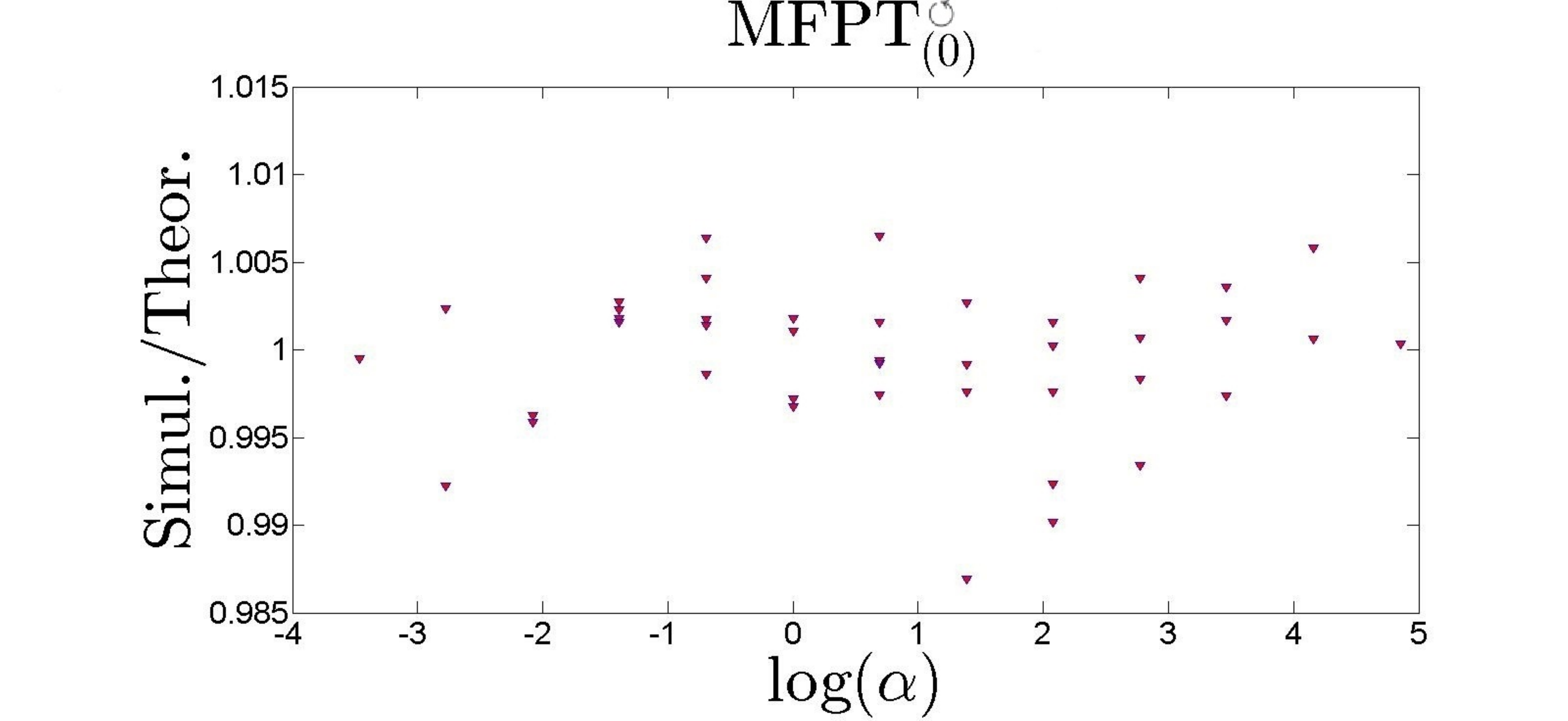}
\protect\caption{\label{fig: MTT Dorso Open-1} Numerical check of the theoretical predictions of Eq. (\ref{eq: 3) MFPT close}) for $\left(y_{f}\right)=\left(0\right)$. In both figures the results of the simulations are divided by the theoretical value of $\mbox{MFPT}_{\left(0\right)}^{\circlearrowleft}$.}
\end{figure}

\begin{figure}[htbp]
\includegraphics[width=\linewidth]{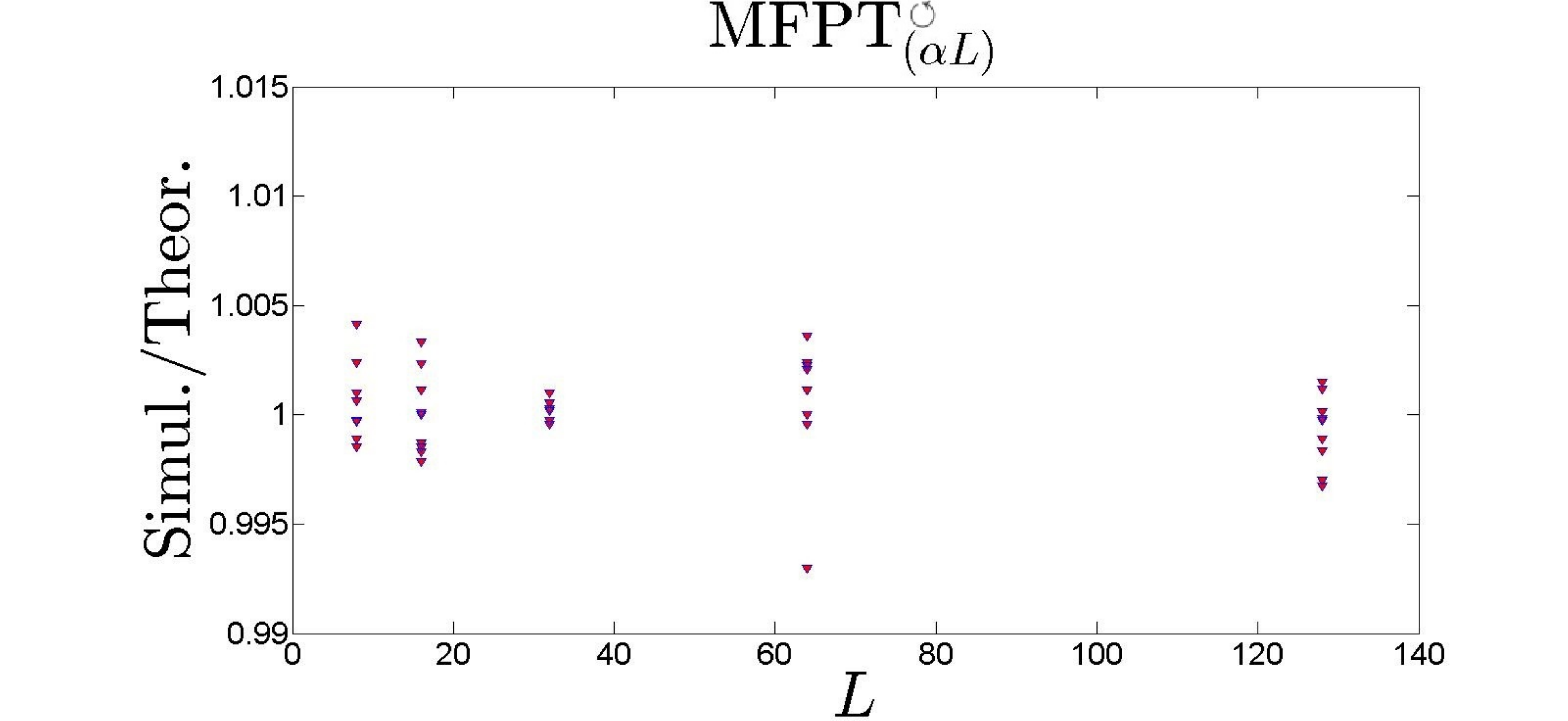}
\includegraphics[width=\linewidth]{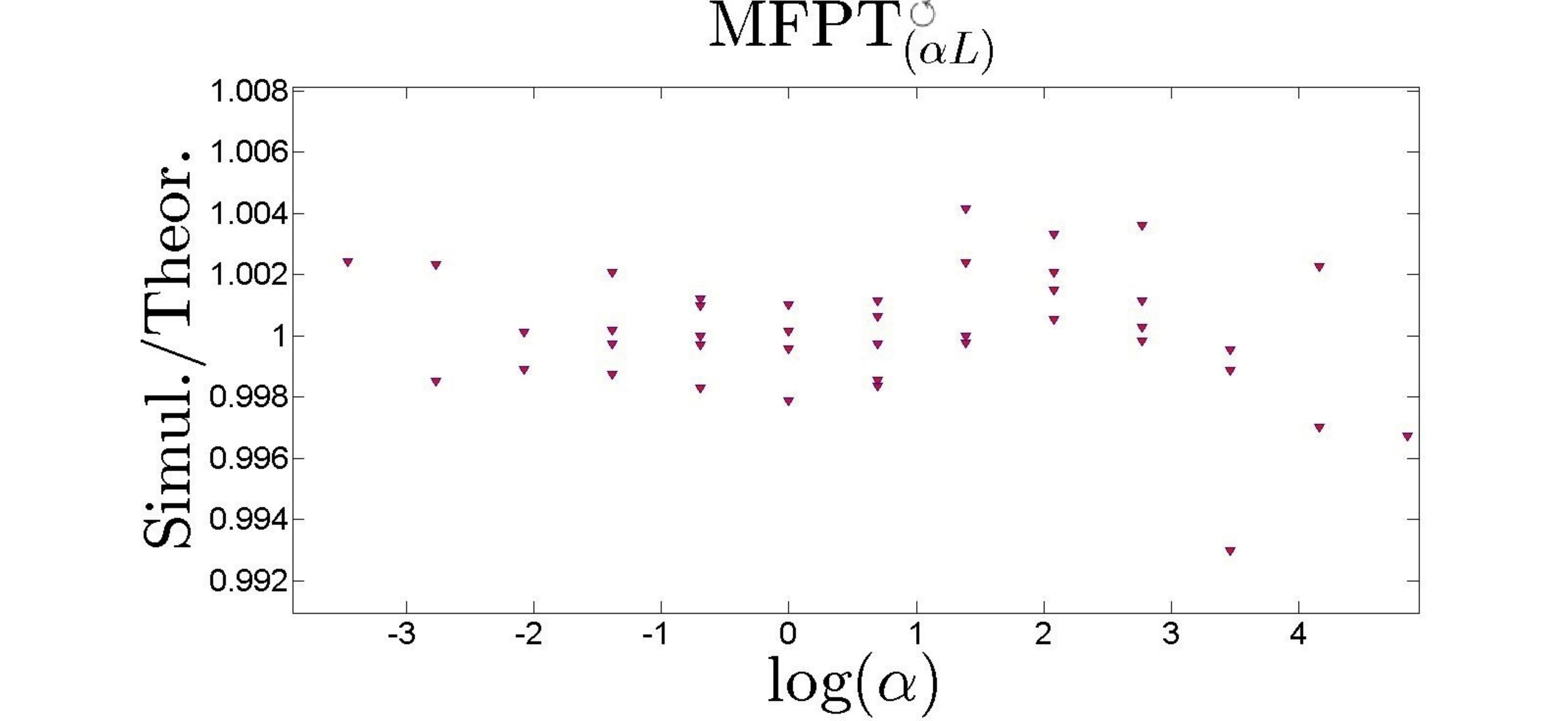}\protect\caption{\label{fig: MTT dente Open-1} Numerical check of the theoretical predictions of Eq. (\ref{eq: 3) MFPT close}) for $\left(y_{f}\right)=\left(\alpha L\right)$. In both figures the results of the simulations are divided by the theoretical value of $\mbox{MFPT}_{\left(\alpha L\right)}^{\circlearrowleft}$.}
\end{figure}

The asimptotic behaviour reads as:
\begin{eqnarray}
\mbox{MFPT}_{f}^{\circlearrowleft}\left(L,\alpha\right) & \sim 	& L^{3}\left\{ \frac{4}{3}\alpha+8\alpha\frac{y}{L}\right\} .\label{eq: 3) AS MFPT Close} \\
														& \sim  & \frac{4}{3}\alpha L^3, \nonumber
\end{eqnarray}
where the last relation holds as longs $y$ is finite.

We have numerically tested this result for two values of $y_{f}$ and for several values of $L$ and $\alpha$, as shown in Figs. (\ref{fig: MTT Dorso Open-1} and \ref{fig: MTT dente Open-1}). In each of those figures the value of $y_{f}$ is fixed and we changed the value of $L$ and $\alpha$. The values of $y_{f}$ are respectively $0$ and $\alpha L$, $L$ takes values from $8$ and $256$, and $\alpha$ goes from $1/32$ to $128$.

\subsection{GMFPT}
We define the Global Mean First Passage Time, GMFPT, as the mean Hitting Time averaged on starting and on ending points:
\begin{eqnarray*}
\mbox{GMFPT} 	& = & \mathbb{E}_{i,f}\left[H\left(i,f\right)\right]; \\
				& = & \frac{1}{V^2}\sum_{i\in\mathcal{V}}\sum_{f\in\mathcal{V}}H\left(i,f\right)
\end{eqnarray*}

GMFPT depends qualitatively on the topological properties of the underlying structure and this has been proven from different perspectives. For instance, it is well known that GMFPT is related with the Kirchhoff index \cite{Klein_Randic}, $K=\mathbb{E}_{i,f}\left[R_{i,f}\right]$ (in fact, from Chandra's formula $H\left(i,f\right)+H\left(f,i\right)=2mR_{i,f}$ \cite{Chandra_commute} it follows straightforwardly that $\mbox{GMFPT}=m\langle R \rangle$).

Moreover, Benichou et al. \cite{Benichou_GMFPT} found a very general asymptotic expression reading as:
\begin{equation}
\mbox{GMFPT}_B \sim\begin{cases}
V & d_{w}<d_{f}\\
V\log\left(V\right) & d_{w}=d_{f}\\
V^{d_{w}/d_{f}} & d_{w}>d_{f}
\end{cases}\label{eq: benichou}
\end{equation}
where $d_w$ is the walk dimension and $d_f$ is the fractal dimension. Their definition of global mean first passage time, is slightly different from ours: theirs is averaged over links,
\begin{eqnarray*}
	\mbox{GMFPT}_B=\frac{1}{m}\sum_{i,f\in\mathcal{V}} \frac{H\left( i,f\right)g\left( i\right)g\left(f\right)}{\left( m-g\left(f\right)\right)}
\end{eqnarray*}
while ours is averaged over vertices:
\begin{eqnarray*}
	\mbox{GMFPT}=\frac{1}{V^2}\sum_{i,f\in\mathcal{V}}H\left( i,f\right).
\end{eqnarray*} 
Despite this different definitions they showed (via numerically simulation) that these lead to the same asymptotics. Now, for comb lattice the walk dimension $d_w$ is not unique \cite{Bertacchi_Zucca} and Einstein's relation $\widetilde{d}\,d_w=2d_f$ does not hold directly and one could therefore calculate GMFPT in a more "pedestrian" way.
Now, we report the exact value of $\mbox{GMFP}$ for the combs structures outlined above, and compare their asymptotic behaviors with Eq. (\ref{eq: benichou}).

\subsubsection{Bidimensional Open Combs}
To calculate the GMFPT for open combs, we start from Eq. (\ref{eq: 3) MFPT open}). We recall that $\mbox{MFPT}_f$ can be written as the sum two terms, the former depends on the final vertex, and the latter is exactly the value of $\mbox{MFPT}_{\left(0,0\right)}$, in a such way that in order to obtain the exact value of GMFPT we have to average only on the former; namely
\begin{equation}
	\mbox{GMFPT}=\mathbb{E}\left[f\left(x_f,y_f\right)\right]+\mbox{MFPT}_{\left(0,0\right)}.
\end{equation}
and after some algebra we find:
\begin{eqnarray}
 \mbox{GMFPT} & = & \sum_{x_f=-L}^L \sum_{y_f=-\alpha L}^{\alpha L} \mbox{MFPT}_f  = \label{eq: GMFPT open} 	\\
 & = &  \frac{1}{3\left(1+2L\right)\left(1+2L\alpha\right)}\left[8L^{2}\left(1+2\alpha+\alpha^{2}\right)+\right.\nonumber \\
 &   & \left. +8L^{3}\left(1+8\alpha+8\alpha^{2}+\alpha^{3}\right)\right.+\nonumber \\
 &   & \left.8L^{4}\left(4\alpha+15\alpha^{2}+5\alpha^{3}\right)+16L^{5}\alpha^{2}\left(2+3\alpha\right)\right], \nonumber
\end{eqnarray}
whose leading order is:
\begin{equation}
\mbox{GMFPT}\left(L,\alpha\right)\sim L^{3}\left(\frac{8}{3}\alpha+4\alpha^{2}\right).\label{eq: 3) AS GMFPT}
\end{equation}

\begin{figure}[htbp]
\includegraphics[width=\linewidth]{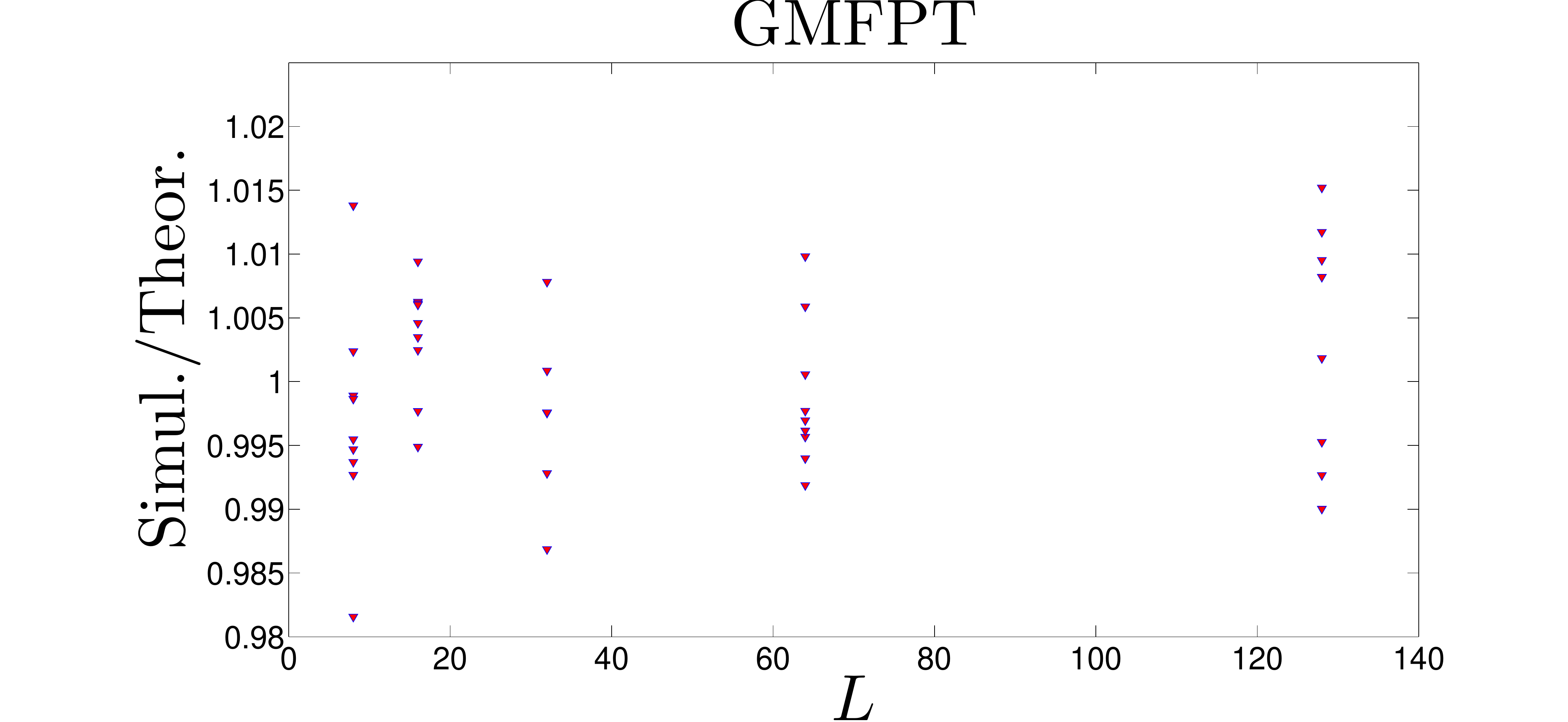}
\includegraphics[width=\linewidth]{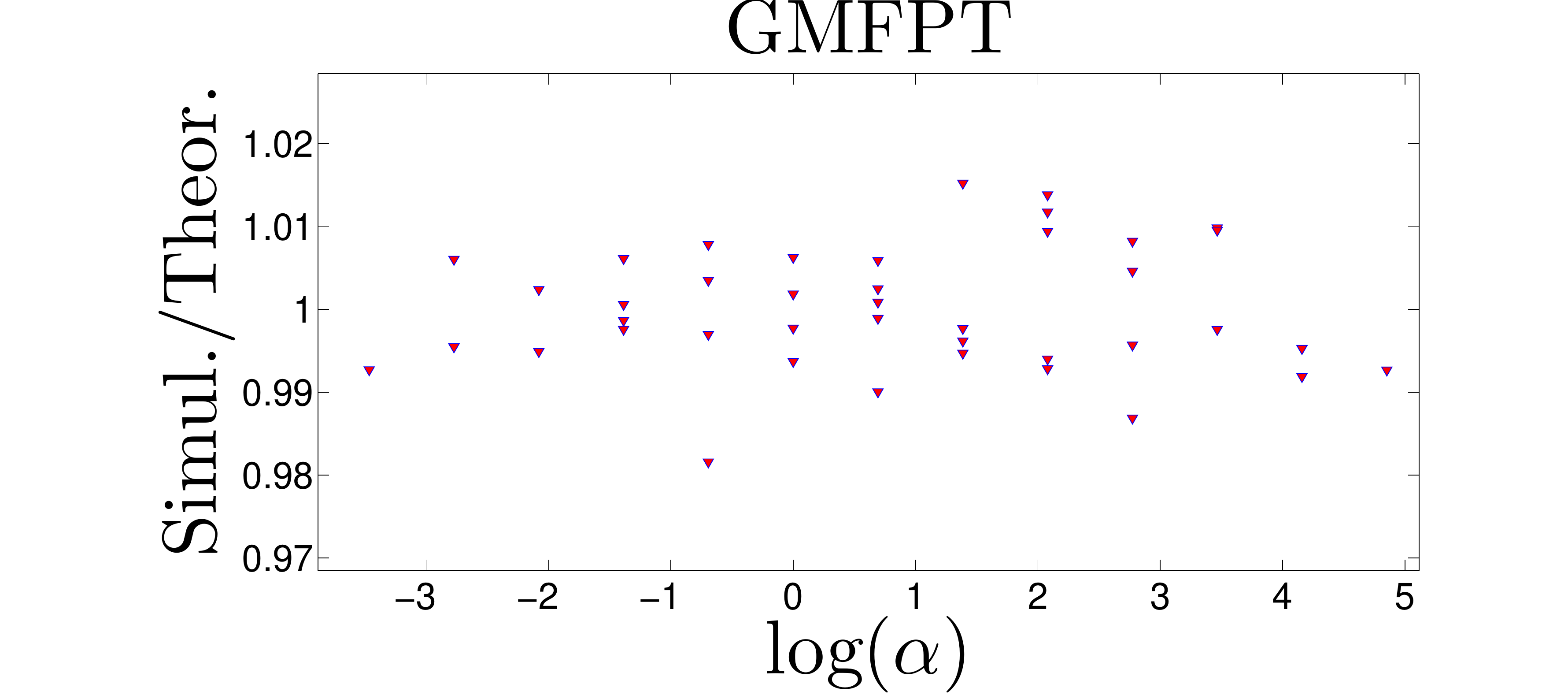}\protect\caption{\label{fig: GMFPT Open} Numerical check of the theoretical predictions of Eq. (\ref{eq: GMFPT open}). In both figures the results of the simulations are divided by the theoretical value of $\mbox{GMFPT}$.}
\end{figure}

We have numerically tested this result for many values of $L$ and $\alpha$, as shown in Fig. \ref{fig: GMFPT Open}. In this figure we changed the value of $L$ from $8$ to $256$ and $\alpha$ from $\frac{1}{32}$ to $128$.

\subsubsection{Bidimensional Close Combs \label{sec: Close}}
We now apply analogous arguments to calculate the exact value of the global mean first passage time for close combs, hereafter referred to as $\mbox{GMFPT}^\circlearrowleft$; averaging $\mbox{MFPT}_f^\circlearrowleft$ over $f$ we find the exact value of $\mbox{GMFPT}^\circlearrowleft$ as
\begin{eqnarray}
\mbox{GMFPT}^{\circlearrowleft} & \left(L,\alpha\right)= & \frac{2}{3}\left[L\left(1+2\alpha\right)\right.+\label{eq: GMFPT close}\\
 &  & \left.+L^{2}\left(1+8\alpha+2\alpha^{2}\right)+L^{3}\left(2\alpha+6\alpha^{2}\right)\right]\nonumber 
\end{eqnarray}
whose leading order is:
\begin{equation}
	\mbox{GMFPT}^{\circlearrowleft}\left(L,\alpha\right)\sim L^{3}\left(\frac{4}{3}\alpha+4\alpha^{2}\right).\label{eq: 3) AS GMFPT Close}
\end{equation}

\begin{figure}[htbp]
\includegraphics[width=\linewidth]{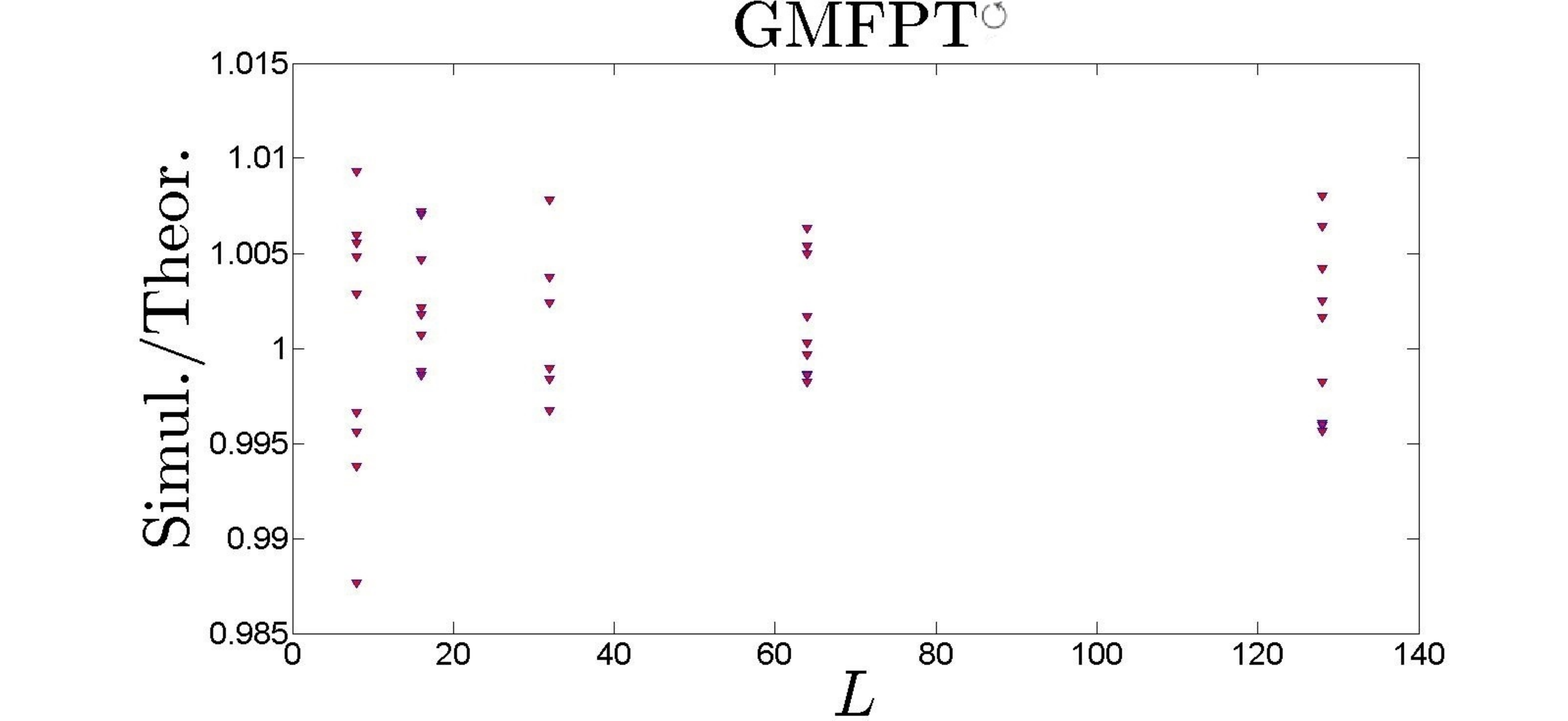}
\includegraphics[width=\linewidth]{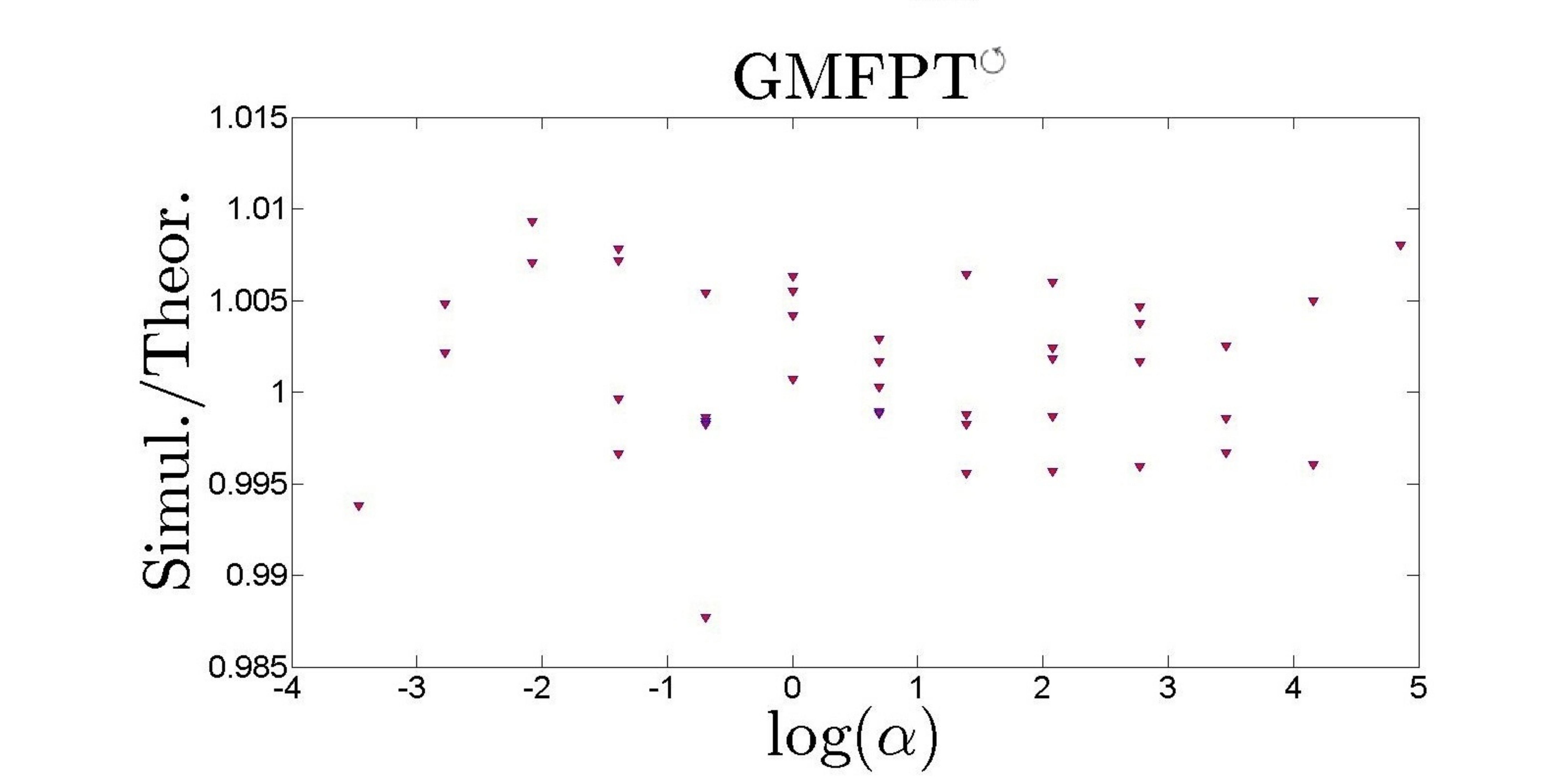}\protect\caption{\label{fig: GMFPT Close} Numerical check of the theoretical predictions of Eq. (\ref{eq: GMFPT close}). In both figures the results of the simulations are divided by the theoretical value of $\mbox{GMFPT}^{\circlearrowleft}$.}
\end{figure}

We have numerically tested this result for many values of $L$ and $\alpha$, as shown in Fig. \ref{fig: GMFPT Close}. In this figure we changed the value of $L$ from $8$ and $256$ and $\alpha$ from $\frac{1}{32}$ and $128$.

Once again we can notice that the coefficient of $\alpha^{2}$ is the same in both combs, but the coefficient of $\alpha^{1}$ is not.
For both GMFPT and GMFPT$^\circlearrowleft$ the leading order $\sim L^3$ is larger than the behaviour expected from a more homogeneous structure with analogous dimensions ($\widetilde{\delta}=3/2$ and $d_f=2$). This underlines once more the peculiar behaviour of combs.

\section{Chemical-Kynetics Perspective}

In this section we frame the previous results within a chemical-kinetics perspective. Imagine having two reactants, say A (static) and B (dynamic). If we are allowed to choose the position of A, but we have no control on B (namely we can not fix its starting point), which will diffuse freely thorought the lattice, we can choose for the static reactant the place which minimize the reaction time. In fact, this is just the node $f$ that minimize $\mbox{MFPT}_{f}$. 

As shown in Sec. [\ref{sec: Open}] and in Sec. [\ref{sec: Close}], the MFPT for both open and close combs can be written as $\mbox{MFPT}_{f}=\phi\left(f\right)+const$. In both combs previously analyzed we found that $\phi\left(f\right)\geq 0$ for every final vertex $f$. In particular, in open combs $\phi\left(f\right)=0$ when $f=\left(0,0\right)$, and in close ones it is zero when $f=\left(i,0\right)$; the maximum value of $\mbox{MFPT}_{f}$ arises when the value of $\phi\left(f\right)$ is maximized, this occurs for $f=\left(\pm L,\pm\alpha L\right)$ for open combs and $f=\left(i,\pm\alpha L\right)$ for close ones. Of course, for close combs there is no dependence on the $x$ coordinate due to the periodic condition on the $x$ axis. 

Now, if we can not control the position of A, but this is stochastic, which is the typical time $\tau_f=\mbox{MFPT}_f$ for the reaction to occur? In particular, the reaction will be considered ``slow'' if $\tau_f>\textrm{GMFPT}$ and ``fast'' if $\tau_f<\textrm{GMFPT}$. 
We characterize the boundary between the two regimes finding those vertices $f$ for which $\mbox{GMFPT}=\mbox{MFPT}_f$. We will obtain this results in the limits of $L\rightarrow\infty$ by imposing the asimptotic equality between Eq. (\ref{eq: 3) AS MFPT}) and (\ref{eq: 3) AS GMFPT}), and between Eq. (\ref{eq: 3) AS MFPT Close}) and (\ref{eq: 3) AS GMFPT Close}). Recalling that $f=\left(x_f,y_f\right)$ we will looking for the functional form $y_f\left(x_f,L,\alpha\right)$ such that  $\mbox{GMFPT}\sim\mbox{MFPT}_{x_f,y_f}$.
\subsection{GMFPT vs MFPT$_{f}$ in Open Combs}
Let us consider the leading value of GMFPT and $\mbox{MFPT}_{f}$ from, Eqs. (\ref{eq: 3) AS MFPT}) and (\ref{eq: 3) AS GMFPT}):
\[
\begin{cases}
\mbox{GMFPT}\left(L,\alpha\right) & \sim L^{3}\left(\frac{8}{3}\alpha+4\alpha^{2}\right)\\
\mbox{MFPT}_{f}\left(L,\alpha\right) & \sim L^{3}\left\{ 4\alpha\left[\frac{1}{3}+X^{2}\right]+8\alpha^{2}|Y|\right\},
\end{cases}
\]
where we normalized the coordinates as $X=x/L$ and $Y=y/\alpha  L$.
Now vertices $f$ for which $\mbox{MFPT}_{f}=\mbox{GMFPT}$ are those whose coordinates fulfill the following equality:
\[
\left(\frac{8}{3}\alpha+4\alpha^{2}\right)=\left[4\alpha\left(\frac{1}{3}+X^{2}\right)+8\alpha^{2}|Y|\right].
\]
Solving for $Y$ we find
\[
|Y|=\left(\frac{1}{2}+\frac{1}{6\alpha}\right)-\frac{X^{2}}{2\alpha}.
\]
It is interesting to note that the fraction of vertices for which the reaction is slow, $F_{\textrm{slow}}$, or fast, $F_{\textrm{fast}}$, are exactly the same for every value of $\alpha$. This is obtained integrating the value of $Y$:
\[
F_{\textrm{slow}}=\intop_{0}^{1}|Y\left(X,\alpha\right)|dX=\frac{1}{2}.
\]
\begin{figure}[htbp]
\includegraphics[width=0.5\linewidth]{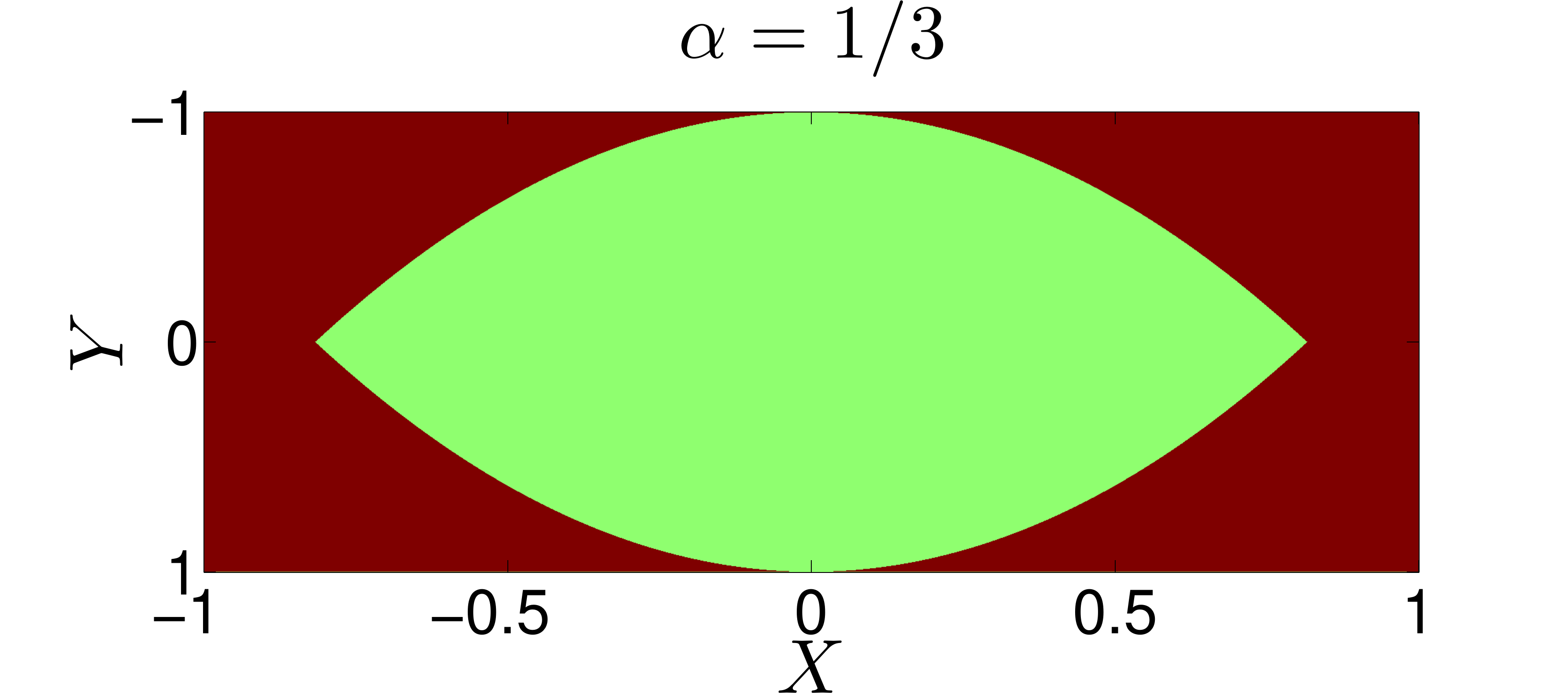}\includegraphics[width=0.5\linewidth]{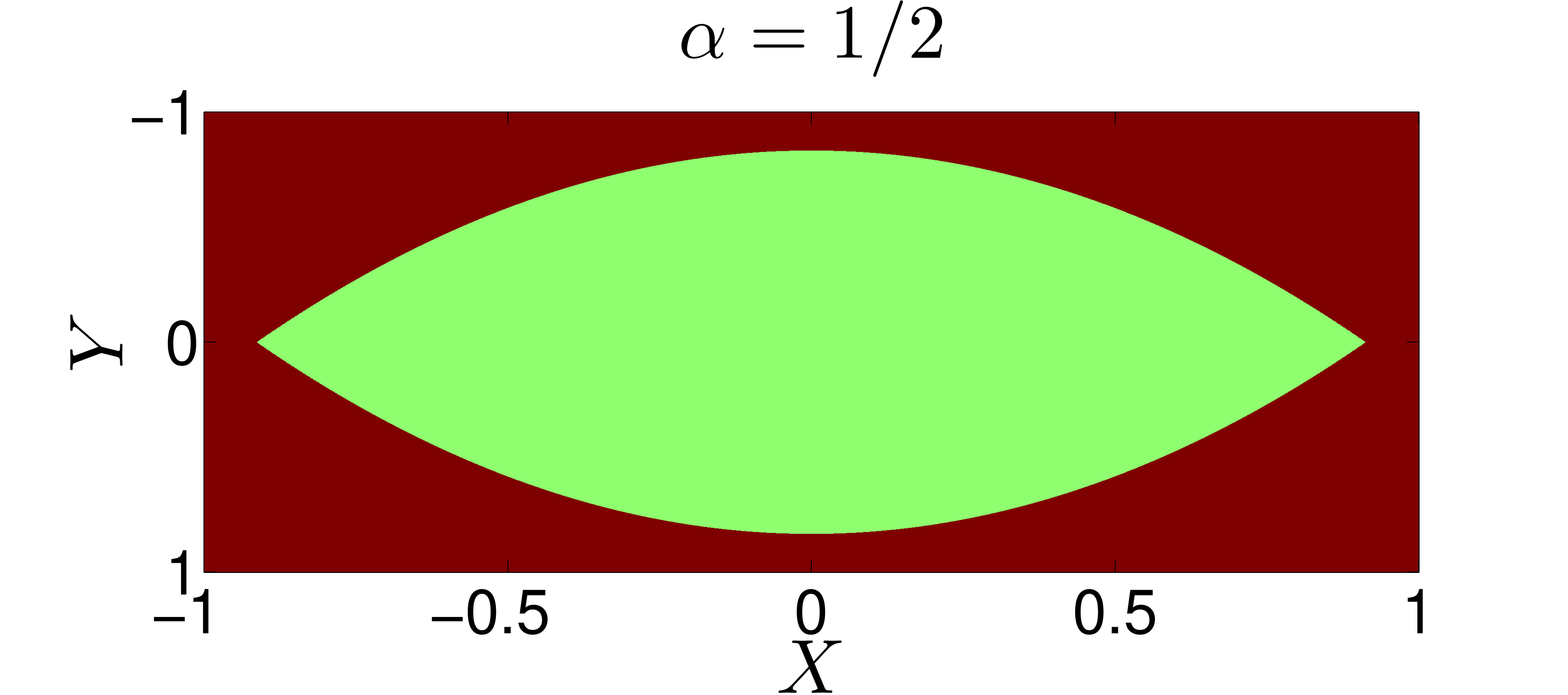}
\includegraphics[width=0.5\linewidth]{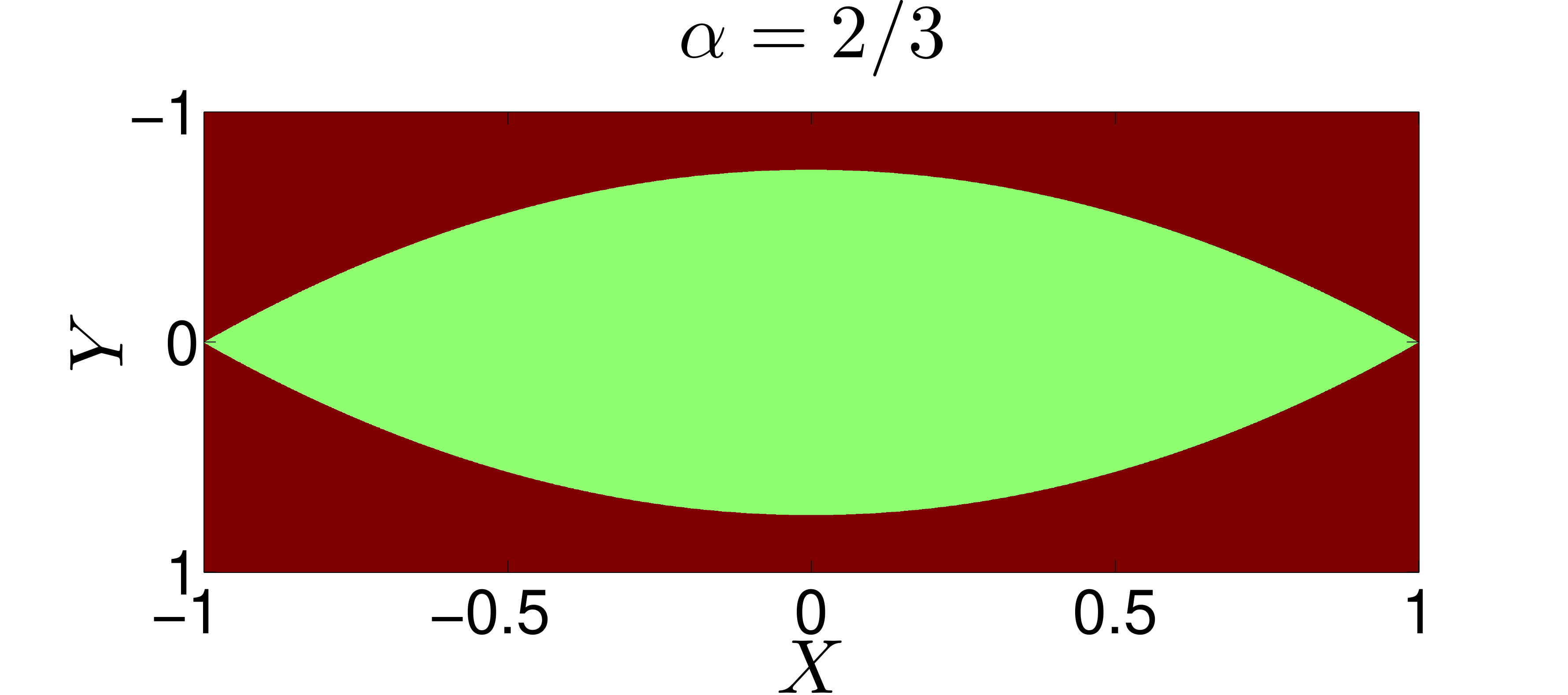}\includegraphics[width=0.5\linewidth]{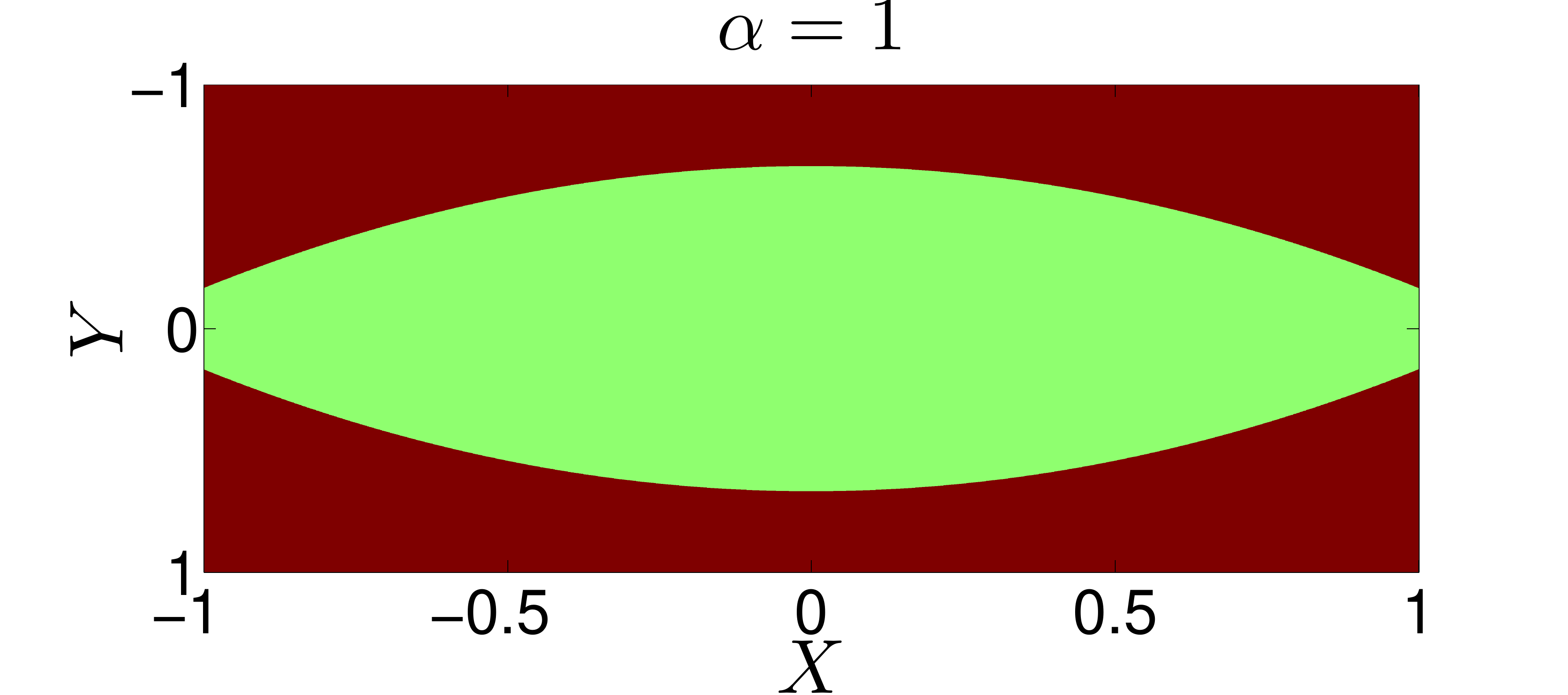}
\caption{\label{fig:Diagramma Open}Phase diagram of $\mbox{MFPT}_{\left(Y\right)}-\mbox{GMFPT}$; those vertices where $\mbox{MFPT}_{\left(Y\right)}>\mbox{GMFPT}$ are colored red, and those where $\mbox{MFPT}_{\left(Y\right)}<\mbox{GMFPT}$ are green}
\end{figure}
The phase diagram is shown in Fig.~\ref{fig:Diagramma Open}.
As we can see, varying the value of $\alpha$ there are three possible scenarios: when $\alpha$ is sufficently small there are some teeth totally composed by ''fast'' vertices (like in $\alpha=\frac{1}{3}$), when $\alpha$ is large enough there are no teeth that are completely ''fast'' or ''slow'' (like in $\alpha=\frac{2}{3}$ and in $\alpha=1$), and an intermediate case where there exist some teeth entirely composed by ''slow'' vertices but no one that is totally composed by ''fast'' vertices (like in $\alpha=\frac{1}{2}$);
\begin{itemize}
\item the first case happens when $|Y|\geq 1$ for $X=0$. This means that $|Y|=\left(\frac{1}{2}+\frac{1}{6\alpha}\right)\geq 1$, and solving in $\alpha$ we find $$\alpha\leq\frac{1}{3};$$
\item the second case happens when $|Y|\geq 0$ for $X=1$. This means that $|Y|=\left(\frac{1}{2}+\frac{1}{6\alpha}\right)-\frac{1}{2\alpha}\geq 0$, and solving in $\alpha$ we find $$\alpha\geq\frac{2}{3};$$
\item the last case happens in the intermediate case: $$\frac{2}{3}>\alpha>\frac{1}{3}.$$
\end{itemize}

\subsection{GMFPT vs MFPT$_{f}$ in Close Combs}

Let us consider the leading terms of GMFPT and $\mbox{MFPT}_f$, Eqs. (\ref{eq: 3) AS MFPT Close}) and (\ref{eq: 3) AS GMFPT Close})
\[
\begin{cases}
\mbox{GMFPT}^{\circlearrowleft}\left(L,\alpha\right) & \sim L^{3}\left(\frac{4}{3}\alpha+4\alpha^{2}\right)\\
\mbox{MFPT}_{f}^{\circlearrowleft}\left(L,\alpha\right) & \sim L^{3}\left\{ \frac{4}{3}\alpha+8\alpha^{2}|Y|\right\},
\end{cases}
\]
the set of vertices $f$ for which $\mbox{MFPT}_{f}\sim\mbox{GMFPT}$ are those whose coordinates fulfill the following equality:
\[
\left(\frac{4}{3}\alpha+4\alpha^{2}\right)=\left\{ \frac{4}{3}\alpha+8\alpha^{2}|Y|\right\}.
\]
Solving for $Y$ we find
\[
|Y|=\frac{1}{2}.
\]
\begin{figure}[tbp]
\includegraphics[width=0.5\linewidth]{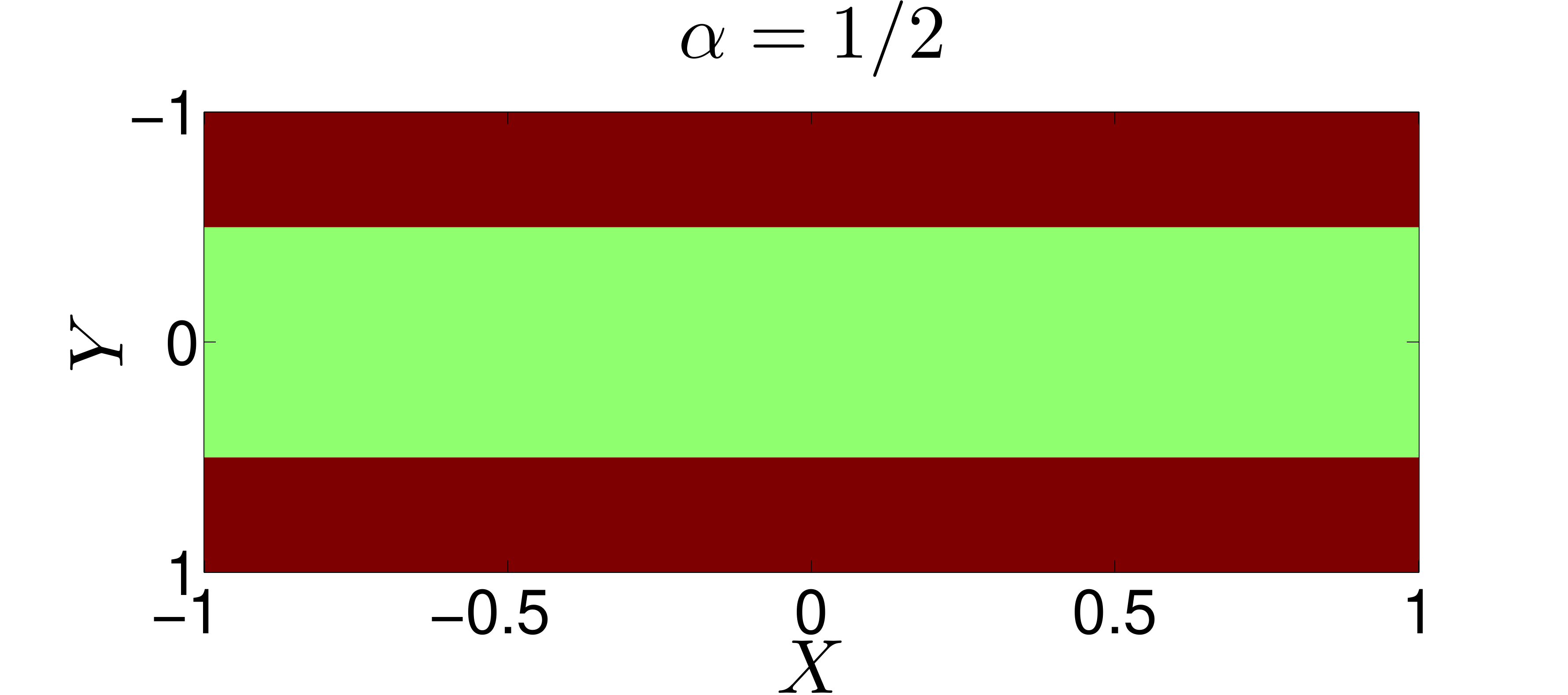}\includegraphics[width=0.5\linewidth]{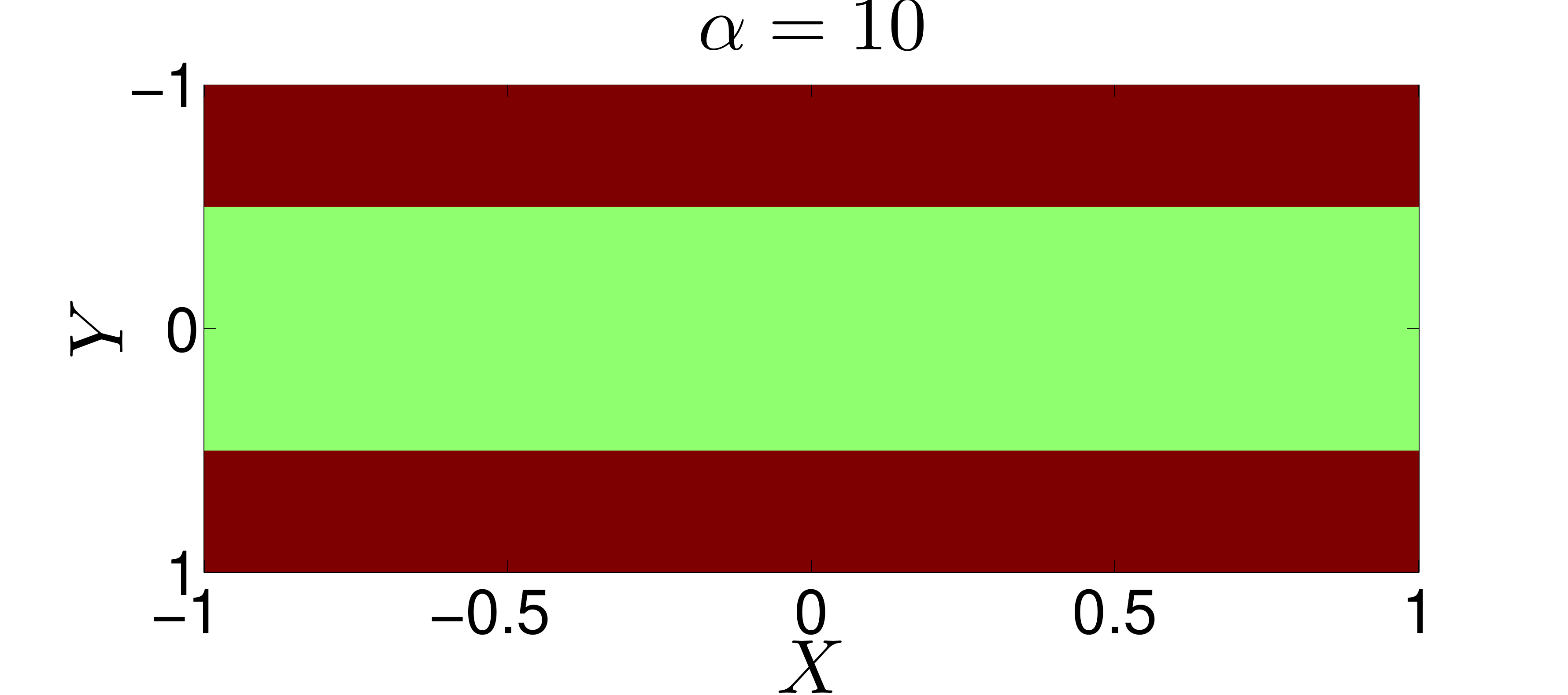}\protect\caption{\label{fig:Diagramma Close} Phase diagram of $\mbox{MFPT}_{\left(Y\right)}^{\circlearrowleft}-\mbox{GMFPT}^{\circlearrowleft}$; those vertices where $\mbox{MFPT}_{\left(Y\right)}>\mbox{GMFPT}$ are colored red, and those where $\mbox{MFPT}_{\left(Y\right)}<\mbox{GMFPT}$ are colored green}
\end{figure}

This means that the location of nodes $f$ such that $\mbox{GMFPT}=\mbox{MFPT}_f$, does not depend neither on $X$ (and this is obvious due to the periodic boundary condition) nor on $\alpha$. Therefore, if the target is placed sufficiently far from the backbone (i.e. at least at half the height of a thoot) the reaction turns out to be slow. Once again the fraction of vertices giving rise to slow and fast reactions is the same. The phase diagram is shown in Fig. \ref{fig:Diagramma Close}.

\section{Conclusions and Future Perspectives}

In this work we found an useful formula, (\ref{eq: H ramificato}), to calculate the exact value of the hitting time, that is an alternative form of Tetali's equation. This formula is particularly useful when the graph has ramifications. In particular we have calculated explicitly the value of $H\left(i,f\right)$ for the bidimensional combs and its leading value for the $d-$dimensional ones. 

We then used these results to calculate the MFPT$_{f}$ and the GMFP for $2$-dimensional combs, observing that the leading value is composed by two terms, one proportional to $\alpha L^{3}$, and the other one to $\alpha^{2}L^{3}$, being $\alpha$ the ratio between the size of the backbone and of the teeth.

As for the GMFPT we noticed that its asymptotic behaviour does not directly depends on its fractal and spectral dimension (as for homogeneous lattices and fractals) and we calculated it exactly. Finally we discussed our results in the context of reaction-kinetics. Infact the MFPT can be seen as the mean time taken by a mobile reactant A to reach a static reactant B, when the starting point of the mobile reactant is not known. Interestingly the typical time for the reaction to occur can be either larger (slow reaction) or smaller (fast reaction) than the GMFPT, and we outlined the set of trap locations for which these regimes are recovered.

\section{Acknowledgments}
EA is grateful to GNFM and Spienza Università di Roma for financial support.

\end{document}